\documentclass[usenatbib]{mnras}
\usepackage[T1]{fontenc}
\usepackage{ae,aecompl}
\usepackage{graphicx}
\usepackage{hyperref}
\usepackage{amsmath}
\usepackage{amssymb}
\usepackage{mathtools}
\usepackage{lscape}
\usepackage{bm}
\usepackage{cleveref}
\usepackage{comment}
\usepackage{bm}

\AtBeginShipout{ %
	\ifnum\value{page}>1 %
	\typeout{* Additional boxing of page `\thepage'} %
	\setbox\AtBeginShipoutBox=\hbox{\copy\AtBeginShipoutBox} %
	\fi %
} %

\def\n{\ensuremath{\mathrm{n}}}
\def\p{\ensuremath{\mathrm{p}}}
\def\L{\ensuremath{\mathrm{L}}}
\def\x{\ensuremath{\mathrm{x}}}
\def\y{\ensuremath{\mathrm{y}}}
\def\vb{\ensuremath{{\mathbf{v}}}}          % velocity
\def\wn{\ensuremath{\boldsymbol{\omega}_n}} % vorticity
\def\Wn{\ensuremath{\hat{\boldsymbol{\omega}}_n}} % vorticity versor
\def\pn{\ensuremath{\mathbf{p}_\n}}          % momentum
\def\hbk{\ensuremath{\hat{\boldsymbol{\kappa}}}}

\def\DR{\ensuremath{\mathcal{R}}}  %Drag parameter
\def\Bc{\ensuremath{\mathcal{B}}}

\def\Om{\Omega}

\title[The effect of non-linear mutual friction] % short title
{The effect of non-linear mutual friction on pulsar glitch sizes and
  rise times } % long title

\author[Celora et al.]{
T. Celora$^{1}$\thanks{T.Celora@soton.ac.uk},
V. Khomenko$^{2}$,
M. Antonelli$^{2}$,
\thanks{mantonelli@camk.edu.pl},
B. Haskell$^{2}$\thanks{bhaskell@camk.edu.pl}
\\
$^1$Mathematical Sciences and STAG Research Centre, University of Southampton, Southampton SO17 1BJ, UK \\
$^{2}$Nicolaus Copernicus Astronomical Center of the Polish Academy of Sciences, ul. Bartycka 18, 00-716 Warsaw, Poland 
}

\begin{document}
\label{firstpage}
\pagerange{\pageref{firstpage}--\pageref{lastpage}}
\maketitle

\begin{abstract}

Observations of pulsar glitches have the potential to provide constraints on the  dynamics of the high density interior of neutron stars. 
However, to do so, realistic glitch models must be constructed and compared to the data. 
We take a step towards this goal by testing non-linear models for the mutual 
friction force, which is responsible for the exchange of angular momentum between the neutron superfluid and the observable normal component in a glitch. 
In particular, we consider a non-linear dependence of the drag force on the relative velocity between superfluid vortices and the normal component, in which the contributions of both kelvin and phonon excitations are included. 
This non-linear model produces qualitatively new features, and is able to reproduce the observed bimodal distribution of glitch sizes in the pulsar population.
The model also suggests that the differences in size distributions in individual pulsars may be
due to the glitches being triggered in regions with different pinning strengths, as stronger 
pinning leads to higher vortex velocities and a qualitatively different mutual friction coupling with respect to the weak pinning case. Glitches in pulsars that appear to glitch quasi-periodically with similar sizes may thus be due to the same mechanisms as smaller events in pulsars that have no preferred glitch size, but simply originate in stronger pinning regions, possibly in the core of the star.

\end{abstract}


%%%%%%%%%%%%%%%%%%%%%%%%%%%%%%%%%%%%%%%%%%%%%%%%%%%%
%  %  %  %  %  %  %  %  %  %  %  %  %  %  %  %  %  %  
%%%%%%%%%%%%%%%%%%%%%%%%%%%%%%%%%%%%%%%%%%%%%%%%%%%%

\section{Introduction}

Neutron stars (NSs) are promising environments in which to study
physics in extreme conditions, and a significant amount of work has been devoted to using
electromagnetic \citep[see e.g.][]{Xray_EOS} and gravitational observations \citep{ligo_EOS} to constrain the equation of state (EOS) of dense matter in the interior of these objects.  
These studies are complementary to those that can be carried out with
terrestrial experiments, as particle accelerators and heavy ion
colliders cannot probe the high density and low temperature behaviour
of the fundamental interactions, nor study the behaviour of matter
with the large isospin asymmetries that characterize NS interiors \citep{NS1book}. In
particular the neutrons are expected to be superfluid at such densities and temperatures \citep{haskell_super,chamel_super}.

Pulsar glitches (sudden spin-ups observed in otherwise spinning down pulsars) are thought to represent a probe into the behaviour of the superfluid interior. 
Most models, in fact, assume that vortices in the superfluid are pinned \citep{ANDERSON1975}, either in the crust or core of the star, and that their sudden unpinning leads to rapid transfer of angular momentum and observed glitch \citep[see e.g. ][]{haskell_review}. 
In these models the dissipative interaction between the superfluid and normal (observable) fluid is given by the so-called `mutual friction' \citep{HV56, BK, langlois_etal98,andersson_MF}, which is mediated by the interaction between vortices and the normal component of the star and sets the observed coupling timescale \citep{alpar_sauls1988}.

Recent observations of glitches in Crab \citep{lyneCrab2015,shaw2018crab}  
and Vela \citep{palfreyman2018Natur} pulsars have been used to obtain constraints on the
mutual friction coefficients \citep{haskell2018corecrust,ashton2019Nat}, which
can shed light on the microscopic interaction occurring in the star
and, ultimately, on the physical region the glitch is triggered. 
This is an interesting point, as calculations of entrainment parameters in the crust show that these may be very large \citep{Carter2006,chamel12}, i.e. there may be very few conduction neutrons in the crust, reducing the amount of angular momentum that can be exchanged and challenging a crustal interpretation of glitches \citep{Andersson2012,Chamel2012a,Delsate:2016}. 
In this case, part of the core superfluid, in which neutron vortices can pin to proton flux
tubes \citep{Muslimov1985,Srinivasan1990,Ruderman1998,alpar_flux_pinn} may be involved, and glitch observations could be used to constrain superfluid gap models, pinning forces and the EOS \citep{Ho2015,pizzochero17,montoli_eos_2020}.

Furthermore, the distribution of glitch sizes appears to be bimodal
\citep{fuentes17}, and some pulsars, such as Vela and the X-ray pulsar
J0537-6910 appear to have mostly large glitches that occur quasi
periodically \citep{AntonopoulouJ05376910}, as opposed to most other pulsars for which the size
distribution is consistent with a power law and the waiting time
distribution with an exponential \citep{melatos07,HowittApJ2018}. This
difference is also likely to be the hallmark of different physical
regimes of the process \citep{fulgenzi2017} and possibly due
to glitches originating in different physical regions of the star
\citep{HaskellDanai2013}. 

A careful determination of mutual friction parameters in the core and
crust of the star, and a study of the system's hydrodynamical response
is crucial in order to make theoretical predictions. In the core
of the star mutual friction is thought to be mainly due to electron
scattering off magnetised vortex cores \citep{alpar84rapid}, while in the
crust energy is dissipated mainly by phonon excitations of the lattice
\citep{Jones1990} and kelvin excitations of the vortices themselves
\citep{epstein_baym92,Jones1992}, which can also occur when vortices
cross flux tubes in the core \citep{Ruderman1998,Link2003L,glampedakis11}. Significant
uncertainties still remain in the determination of crustal mutual
friction parameters, which in turn impact on predictions of the
glitch rise time and observed size \citep{HP12,antonelli17,SC16}. 
A recent step forward was taken by \cite{Graber2018} who, following the approaches of \cite{epstein_baym92} and \cite{Jones1992}, calculated the kelvin friction parameters at different densities in the crust and compared the results to Vela glitches.

In this paper we include an additional ingredient in the calculation
which, we show, has a strong impact, both qualitative and quantitative,
on the glitch features, namely the dependence of the mutual friction
parameters on the relative velocity between superfluid vortices and
the normal component. 
This issue is of fundamental importance in the crust, as kelvin mutual friction
depends strongly on the relative velocity \citep{epstein_baym92,Jones1992} and is
suppressed for low velocities, at which phonon contributions dominate
\citep{Jones1990,jones1990temperature}. If vortices unpin and initially move at large
velocities, they will thus experience a varying drag as the system
relaxes towards equilibrium.
The problem is, however, of more general relevance, as relative
motions between the superfluid and normal components may lead to
turbulence \citep{andersson_turbulence,peralta2006} and also to a different
velocity dependence of the mutual friction parameters, as is, for
example, well known for objects falling through the
atmosphere on Earth, for which the terminal velocity can be obtained
by considering a drag that scales with the square of the velocity,
rather than linearly as would be expected if the air flow were laminar. 

In the following we first examine the problem of mutual
friction in the case of velocity dependent parameters, and discuss the
different possible physical regimes. We then move on to
discuss the mutual friction in the crust of a neutron star,
and present a model which includes both kelvin
and phonon contributions. We show that this model predicts a
qualitatively different rise than standard mutual friction
models, which is consistent with recent observations of a
glitch in the Vela pulsar \citep{palfreyman2018Natur,ashton2019Nat}.
We also apply the model to a population of glitching pulsars and show that it produces a bimodal glitch size population that has an excess of large glitches. 
  
%%%%%%%%%%%%%%%%%%%%%%%%%%%%%%%%%%%%%%%%%%%%%%%%%%%%
%  %  %  %  %  %  %  %  %  %  %  %  %  %  %  %  %  %  
%%%%%%%%%%%%%%%%%%%%%%%%%%%%%%%%%%%%%%%%%%%%%%%%%%%%

\section{Mutual friction: linear Drag in absence of turbulence}

Let us begin our analysis by reviewing the standard derivation of the
mutual friction force in neutron stars \citep{mendell1991dissipative,carter2005III,andersson_MF}. 
Following \cite{PC02}, we deal with a system composed of a charge neutral mixture of protons and electrons and everything coupled to it on short timescales ( so that we can treat them as a single  component, the `normal' one hereafter dubbed ``p'') plus a superfluid neutron component (hereafter dubbed ``n'') which, due to
the stellar rotation, is threaded by an array of quantized vortices.

In the absence of interactions between the superfluid and the normal
component, the vortex line velocity $\vb_{\L}$ is equal to the
bulk velocity $\vb_{\n}$ of the neutrons (where both  $\vb_{\L}$ and  $\vb_{\n}$ are orthogonal to the vortex line, see e.g. \citealt{DonnBook1991}). 
This result can be understood by analysing the forces acting on
a vortex, which to a very good degree of approximation can be treated as a
massless object \citep{baym1983,sonin87,DonnBook1991}, 
although see \cite{simulaMass} for a recent discussion on the effective mass of a quantum vortex. 
In this case the only force acting on a free vortex line is the Magnus
force $\mathbf{f}_M$ per unit length of vortex, 
\begin{equation}
f_M^i = \rho_\n \varepsilon^{ijk}\kappa_j( v_k^\L - v_k^\n ) \;,
\end{equation}
where $\rho_\n$ is the density of neutrons and $\bm\kappa$ is a vector
aligned locally with the vorticity, such that 
$\bm\kappa = \kappa \, \hat{\bm\kappa}$ and $\kappa=h/2m_\n$ is the quantum of circulation. 
The requirement of force balance leads to $\mathbf{f}_M=0$ and thus $\vb_{\L}=\vb_{\n}$. 
In the more general case in which  the vortices interact also with the normal component flowing with velocity $\vb_{\p}$, the vortex velocity $\vb_{\L}$ will in general differ from both $\vb_\n$ and $\vb_\p$.

If we describe the dissipative interaction between the vortices and the normal
component in terms of a linear drag force, the equation of motion of a single vortex line is 
\begin{equation}\label{eq:VortexDynamics}
    f_M^i + f_D^i = 0 \; , 
\end{equation}
with the drag force per unit length defined as:
\begin{equation}
f_D^i = - \eta (v_\L^i - v_\p^i) \;.
\end{equation}
The drag coefficient $\eta$ is related to the dimensionless quantity
$\DR$ often used in the literature via $\DR  = \eta / \kappa\rho_\n $. It is useful to work in the normal component rest frame where the  \cref{eq:VortexDynamics} reads 
\begin{equation}
    \hbk \times \big(\vb_{\L\p} - \vb_{\n\p}\big) - \DR \vb_{\L\p} = 0 \,.
\end{equation}
We can solve this equation to write the vortex velocity $\vb_{\L\p}=\vb_\L-\vb_\p$ in terms 
of the lag $\vb_{\n\p}=\vb_\n-\vb_\p$, namely
\begin{equation}\label{eq:LinearVortexVelocity}
\vb_{\L\p} 
= 
-\frac{\DR}{1+\DR^2} \, \hbk \times\vb_{\n\p} 
- 
\frac{1}{1+\DR ^2} \, \hbk \times \big( \hbk \times \vb_{\n\p} \big) \, .
\end{equation}
In the limit of ``weak drag'', $\DR \ll 1$, one can neglect the first term 
to get $\vb_{\L\p} \approx - \hbk\,\times (\hbk\,\times\,\vb_{\n\p} )$, as
expected for the no-drag case. On the other hand, if we consider the
``super strong drag'' regime, that is $\DR \gg 1$, the previous equation
gives $\vb_{\L\p} \approx 0$, implying that the vortex is effectively pinned.

Now, the force per unit volume acting between the normal component and the
superfluid is found by averaging over all the vortices in the fluid element, or rather over the total length of vortex lines $L$ in the element. In general this depends on the
nature of the flow, as a turbulent flow can tangle the vortices and increase the length of vortex in an element \citep{vinen1957III,schwarz88}. 
We will discuss this possibility in the following sections. In the case of an array of straight and aligned vortices, it is possible to introduce the vortex density $n_v$ on a unit surface orthogonal to $\hbk$ and use it to get the averaged force per volume element. 
Then, using Newton's third law, the force per unit volume between the vortices and the
fluids can be included in the hydrodynamical equations (the full form will be given in \eqref{totalEUL}) as
\begin{equation}
\begin{split}
\rho_n \partial_t \vb_n + \dots &= - n_v \, \textbf{f}_M \\
\rho_p \partial_t\vb_p + \dots &= -n_v\,\textbf{f}_D \,.
\end{split}
\label{eq:MF1}
\end{equation}
From \cref{eq:MF1} we can read the mutual friction force $\textbf{F}_{MF}$, defined as the force
exerted by the normal component on the superfluid, namely
\begin{equation}
	F^i_{MF} \, = \, - n_v \, f^i_M \,  =  \, n_v\,f^i_D\, ,
\end{equation}
where the vortex density $n_v$ can be seen as a measure of the macroscopic vorticity $\omega^i_\n$ vector via
\begin{equation}
	\omega^i_\n = m_\n^{-1}\, \varepsilon^{ijk}\,\partial_j\,p^\n_k = \kappa\,n_v\,\hat\kappa^i \, .
	\label{crespelle}
\end{equation}
Therefore, the mutual friction force can be written in such a way that only macroscopic hydrodynamical quantities appear \citep{andersson_MF}
\begin{equation}
	\textbf{F}_{MF} = \rho_\n\Big( \Bc_c\, \wn \times \vb_{\n\p} + \Bc_d\,\Wn \times \big(\wn \times \vb_{\n\p}\big)\Big) \,,
\end{equation}
where $\Wn = \hat{\boldsymbol{\kappa}}$ and 
\begin{equation}
\Bc_c = \frac{\DR^2}{1+\DR^2} 
\qquad \qquad 
\Bc_d =\frac{\DR}{1+ \DR^2} \, .
\label{tonno}
\end{equation}
Therefore, the mutual friction force is composed of a Coriolis-like part which  is proportional to $\Bc_c$, and a dissipative part proportional to $\Bc_d$.

%%%%%%%%%%%%%%%%%%%%%%%%%%%%%%%%%%%%%%%%%%%%%%%%%%%%
%  %  %  %  %  %  %  %  %  %  %  %  %  %  %  %  %  %  
%%%%%%%%%%%%%%%%%%%%%%%%%%%%%%%%%%%%%%%%%%%%%%%%%%%%

\section{Vortex motion with non-linear drag}

In this section we consider the more general case of a non-linear drag
force. In the standard picture, presented in the previous section, the
drag parameter $\DR$ is taken to be a constant, which can be estimated
via microphysical calculations of energy dissipation rates in specific
channels. For instance, in the NS crust, for some
relatively high values of the vortex velocity $v_{\L\p} \gtrapprox
10^{4} \text{ cm s}^{-1}$, energy is dissipated mainly by Kelvin waves
propagating along the vortex line \citep{epstein_baym92}. On the
other hand, for lower vortex velocities the Kelvin waves are
suppressed and energy is dissipated via
excitation of phonons  in the crustal lattice
\citep{Jones1990}. Hence, the drag parameter $\DR$
itself depends on $|v_{\L\p}|$ \citep[see][]{Jones1992}. As a result, we are interested in dealing with the more general case in which  the additional dependence of the drag force on the relative speed $|\vb_{\L\p}|$  is encoded in $\tilde \DR$,
\begin{equation}
    f_D^i = - \rho_\n\,\kappa\, \tilde\DR \, v_{\L\p}^i 
    \; , \qquad
    \tilde \DR = \tilde \DR (|\vb_{\L\p}|) \, .
\end{equation}
Of course, the linear drag case is recovered once $\tilde \DR$ is assumed to be a constant. 

While the case we are explicitly considering is that of a straight vortex array and a 
velocity dependent drag parameter $\tilde \DR$, so that the mutual friction force 
can be written
\begin{equation}
F^i_{MF}=- n_v \rho_\n\,\kappa\, \tilde\DR \, v_{\L\p}^i 
\label{prezzemolo}
\end{equation}
this is mathematically equivalent to considering a turbulent tangle of vortices, with vortex length per unit volume $\tilde{L}=\tilde{L} (|\vb_{\L\p}|)$. In this latter case, the mutual friction force takes the form 
\begin{equation}
F^i_{T}=- \tilde{L} \rho_\n\,\kappa\, \DR \, v_{\L\p}^{i}\, ,
\end{equation}
which is formally the same as \cref{prezzemolo}.
The conclusions we will obtain in the following for different
functional dependence of $\tilde \DR$ on the vortex line velocity can thus be
directly applied to the case in which $\tilde{L}$ has the same
dependence, so that e.g. the qualitative behaviour of the coupling for
$\tilde{\DR}\propto{|\vb_{\L\p}|^2}$ can be applied to the standard case
of fully developed isotropic quantum turbulence as well, in which $\tilde{L}\propto{|\vb_{\L\p}|^2}$
\citep{vinen1957III,andersson_turbulence}.

To proceed it is useful to introduce the usual cylindrical coordinates system
$(\hat e_x, \hat e_\varphi , \hat e_z)$ where the z-axis is aligned with the vortices, i.e. $\hbk = \hat e_z$. 
We also consider the velocities of the two components to be azimuthal, so that $\vb_{\n\p} = x\,\Omega_{\n\p} \hat e_\varphi$. 
On the timescales of the observable glitch dynamics this is true for a fluid element (at least on average), and allows us to simplify the analysis. However, on shorter timescales vortex accumulation
\citep{khomeko2018PASA} and counterflow along the axis of a vortex due to bending \citep{khomeko2019} may give rise to instabilities leading to turbulence or vortex avalanches.

In components, the force balance \cref{eq:VortexDynamics} gives
\begin{equation}\label{eq:FormalComponentSolution}
\begin{split}
    &v_{\L\p}^\varphi - v_{\n\p}^\varphi + \tilde \DR \,v_{\L\p}^x = 0 \\
     &v_{\L\p}^x - \tilde \DR\,v_{\L\p}^\varphi = 0 \, ,
\end{split}    
\end{equation}
that can be rearranged as 
\begin{equation}
\label{eq:BMF}
    v_{\L\p}^x = \frac{\tilde\DR}{1+\tilde\DR^2} v_{\n\p}^\varphi = \Bc_{MF}\, v_{\n\p}^\varphi \, .
\end{equation}
This expression is only formal since we have that the drag parameter
$\tilde\DR$ still has an implicit dependence on $|\vb_{\L\p}|$. The
solution in \cref{eq:BMF} also defines the mutual friction
  coefficient $\Bc_{MF}$, which is introduced in
such a way that it reduces to $\Bc_d$ in the linear drag case (when $\tilde \DR$ is a constant).

The last equation we need is the azimuthal component of the vortex line velocity,
\begin{equation}\label{eq:BMF2}
v_{\L\p}^\varphi = \frac{v_{\n\p}^\varphi }{1+\tilde\DR^2}.
\end{equation}
Combining \cref{eq:BMF} and \cref{eq:BMF2} we obtain
\begin{equation} \label{modulo}
|\vb_{\L\p}|=\frac{|\vb_{\n\p}|}{\sqrt{1+\tilde{\DR}^2}}
\end{equation}
which, given a functional dependence of $\tilde \DR^2$ on
$|\vb_{\L\p}|$, can be solved to eliminate the vortex line velocity
from the hydrodynamical equations in \cref{eq:MF1}. 
We will analyse specific forms of the mutual friction in the following, but let us note here that \cref{modulo} shows that the difference between $|\vb_{\L\p}|$ and $|\vb_{\n\p}|$ is of $\mathcal{O}(\tilde\DR^2)$, so that to very good approximation one can take $|\vb_{\L\p}|\approx |\vb_{\n\p}|$ for $ \tilde \DR \ll 1$, and simply solve the hydrodynamical equations in \cref{eq:MF1} directly for $\tilde \DR(|\vb_{\L\p}|)\approx \tilde \DR(|\vb_{\n\p}|)$ (or equivalently, if dealing with turbulence, for $\tilde{L} (|\vb_{\L\p}|)\approx \tilde{L} (|\vb_{\n\p}|)$).

\subsection{Power Law drag}\label{sec:PowerLawdrag}

Let us start with a simple prescription, and consider a power law behaviour for the drag force, 
\begin{equation}
	f_D^i = - \eta_\beta\,|\vb_{Lp}|^\beta\,v_{Lp}^i \, ,
	\label{eq:padrepio}
\end{equation}
where the physical dimension of the viscous parameter $\eta_\beta$ depends on the explicit value of $\beta$. 
Despite its simplicity, this prescription is applicable to several physical setups. 
For example, in the presence of classical turbulence one has $\beta=1$ (the standard case of objects moving in a fluid at high Reynolds number, according to Newton's drag law), and for isotropic quantum turbulence $\beta=2$, although polarized turbulence is likely to require the
use of multiple power laws to describe the drag \citep{andersson_turbulence, MongioviJouTurbulence07}. 
Negative values of $\beta$ do not have an hydrodynamical
interpretation, but microphysical calculations of Kelvin drag in the crust \citep{Jones1992,epstein_baym92,Graber2018} and also core of the star if the protons are in a type-II superconducting state \citep{Link2003L,haskellGlampe2014}, suggest that $\beta=-3/2$: this case is of particular interest for NSs and will be considered in detail in the following.

To work with a dimensionless drag parameter $\DR$ we introduce a microscopic parameter $v_0$ (with  the dimension of a velocity) and rewrite the drag force in \cref{eq:padrepio} as 
\begin{equation}\label{eq:manibucate}
	f_D^i = - \kappa\,\rho_n\,\DR \,\Big( \frac{|\vb_{Lp}|}{v_0}\Big)^\beta\, v_{Lp}^i \, ,
\end{equation}
where the constant and dimensionless drag parameter is $\DR = \eta_\beta\,v_0^\beta/\kappa\,\rho_n$. 
In terms of the drag coefficient $\tilde \DR$ previously introduced we have that 
\begin{equation}
	\tilde\DR \, = \, \Big(\frac{|\vb_{\L\p}|}{v_0}\Big)^\beta \, \DR  \, . 
\end{equation}
Now, to solve the equations of motion for the vortex line it is helpful to introduce the dimensionless variables 
\begin{equation}\label{eq:DissipationSine}
	s=\frac{v_{Lp}^\varphi}{|\vb_{\L\p}|} \quad  , \quad \sqrt{1-s^2} = \frac{v_{\L\p}^x}{|\vb_{\L\p}|} 
\end{equation}
and
\begin{equation}
    \chi_L = \frac{|\vb_{Lp}|}{v_0}\quad , \quad \chi_{np} = \frac{|\vb_{np}|}{v_0} 
\end{equation}
so to rewrite \cref{eq:FormalComponentSolution} as 
\begin{equation}
\begin{split}
&s\,\chi_L - \chi_{np} +\DR \,\sqrt{1-s^2}\,\chi_L^{\beta +1} = 0 
\\
&\sqrt{1-s^2} - \DR\,\chi_L^\beta\, s = 0 \,.
\end{split}
\end{equation}
Note that $s=\text{cos}(\theta_D)$, where $\theta_D$ is the `dissipation angle' introduced by \cite{epstein_baym92}, see also \citet{link14}. 
Starting from \cref{eq:FormalComponentSolution} one can  show that 
\begin{equation}
\label{eq:PowerLawAlgebraicEq}
\begin{split}
    &\chi_L = s\,\chi_{np} \\
    &\DR^2\,\chi_{np}^{2\beta}\,s^{2\beta + 2} + s^2 -1 = 0 \, .
\end{split}
\end{equation}
Since $0\leq s\leq 1$, namely $0\leq \theta_D\leq \pi/2$, we see that the lag between the
vortices and the normal component must always be smaller than the lag between the neutron ($\n$) and normal ($\p$) component. It is also interesting to observe that, since $\tilde \DR = \DR\,\chi_{\n\p}^\beta\,s^\beta$, the mutual friction coefficient $\Bc_{MF}$ introduced in \cref{eq:BMF} reads
\begin{equation}
	\Bc_{MF} =  s\,\sqrt{1-s^2} = \frac{1}{2} \text{sin}(2\theta_D) \,.
\end{equation}
Different values of $\beta$ correspond to different phenomenological models for the dynamics of vortex lines and the evolution of the system will depend  on the choice for the index $\beta$.

\begin{figure}
    \centering
    \includegraphics[width = 0.99 \columnwidth]{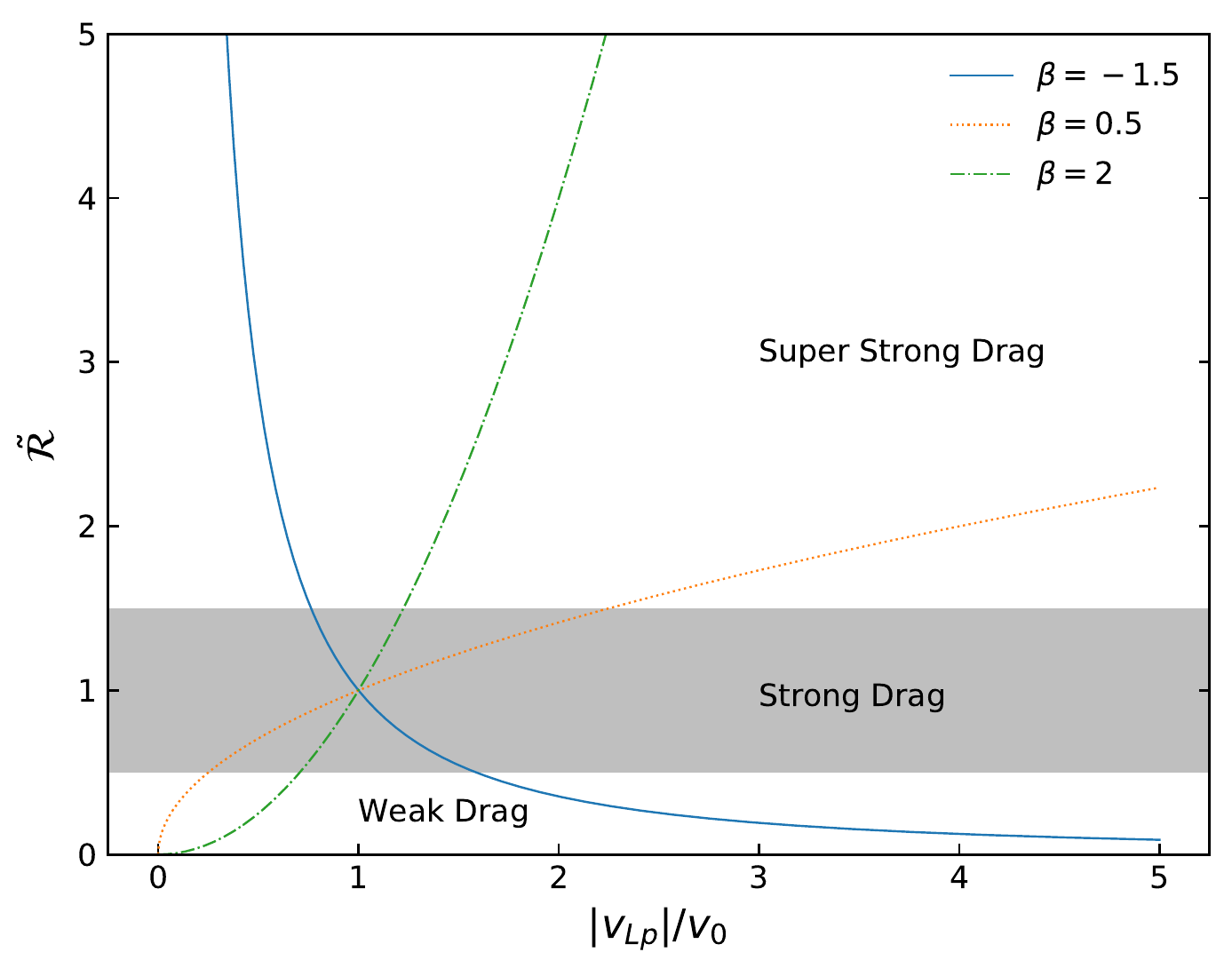}
    \caption{
    Qualitative plot of the drag parameter $\tilde\DR$ as a
      function of $|\vb_{Lp}|/v_0$, for different values of $\beta$ and having chosen a fixed value of $\tilde\DR=1$ for $|\vb_{Lp}|=v_0$. Three regions are highlighted, the `weak drag' region in which $\Bc_{MF}\sim \tilde{\DR}$, the `strong drag' region in which $\Bc_{MF} \lesssim 1/2$ and a `super strong drag' one where the  mutual friction goes to zero as $\Bc_{MF} \sim \tilde{\DR}^{-1}$. Therefore, a drag force described by a power law with a negative $\beta$ can be used to model the effect of initially free vortices repinning.
For positive values of $\beta$ the opposite is true, and the coupling becomes weak as $|\vb_{Lp}|$ decreases. For the case $\beta=2$, which describes mutual friction in an isotropic turbulent tangle, the drag is initially strong, but rapidly decreases as the recoupling proceeds.
}
    \label{fig:BetaTransition}
\end{figure}

To better discuss this point, let us first  remark that it is always possible to choose $v_0$ such that $\DR=1$,  implying $\tilde\DR\approx1$ when $|\vb_{Lp}|\approx v_0$ (this regime corresponds to the largest possible value for the mutual friction parameter, $\Bc_{MF}=1/2$). 
This value of $v_0$ defines three velocity ranges $|\vb_{Lp}| \ll v_0$, $|\vb_{Lp}| \approx v_0$ and $|\vb_{Lp}| \gg v_0$, that are related to three mutual friction regimes. 
This is sketched, for different prescriptions of $\beta$, in \cref{fig:BetaTransition}, where  we show the dependence of the effective drag parameter $\tilde \DR$ on $|\vb_{Lp}|$, having chosen a fixed value of $\tilde\DR=1$ for $|\vb_{Lp}|=v_0$. 

According to \cref{eq:PowerLawAlgebraicEq}, the vortex velocity $|\vb_{Lp}|$  decreases as the two components recouple during the spin-up phase of a glitch, simply because the initial lag is decreasing as well. 

For $\beta<0$ and an initial velocity $|\vb_{Lp}| > v_0$,  the drag increases as the two components recouple and rapidly enters the strong $\tilde \DR \approx 1$ regime, as we can see for the $\beta=-3/2$ case  which is relevant for kelvin mutual friction. 
If during the recoupling process the lag becomes so small that $|\vb_{Lp}| < v_0$, we enter the `super-strong' drag regime, where $\tilde\DR$ diverges. 
In this regime the friction coefficient goes to zero as  $\Bc_{MF} \sim \tilde{\DR}^{-1}$, so that a negative $\beta$ could be used to mimic the repinning process, namely a suppression of the mutual friction.

For $\beta>0$ the opposite is true, and the coupling strength decays  (more or less rapidly depending on the actual value of $\beta$) as the lag decreases. Such a model, for initial conditions such that $|\vb_{Lp}| > v_0$, could be used to describe phenomenologically a situation in which a pinned vortex configuration undergoes unpinning, passes trough a phase of strong drag (in which the recoupling of the components is very fast) and possibly a final part in which $|\vb_{Lp}| \ll v_0$ and the recoupling proceeds with a much slower timescale.   

In fact the power law prescription in \cref{eq:padrepio} can be used to model transitions between different dynamical regimes at different relative vortex velocities. 
If, for example, one has two microscopic estimates $\DR_1$ and $\DR_2$ for the drag parameter in two different regimes such that
\begin{equation}
\begin{split}
	&\tilde\DR (\vb_{Lp} \approx v_1)  \approx \DR_1 = \DR\,\Big(\frac{v_1}{v_0}\Big)^\beta\\
   &\tilde\DR (\vb_{Lp} \approx v_2)  \approx  \DR_2=\DR\,\Big(\frac{v_2}{v_0}\Big)^\beta \; ,
\end{split}
\end{equation}
we may interpolate between the two values using a power law approximation for the drag with exponent
\begin{equation}
	\beta_{12} = \frac{\text{log}(\DR_2/\DR_1)}{\text{log}(v_2/v_1)} \;. 
\end{equation}
The final interpolated drag function reads
\begin{equation}
    \tilde\DR   \, = \, \DR_1  \,\Big(\frac{|\vb_{Lp}|}{v_1}\Big)^{\beta_{12}} 
    \, = \,
    \DR_2  \,\Big(\frac{|\vb_{Lp}|}{v_2}\Big)^{\beta_{12}} \, .
\end{equation}
As a result, the choice of a non-linear drag being a simple power law
may be used for modelling the transition between different effective
drag regimes related to the activation of different dissipation
channels. It can also be used as a simple model for the repinning process: we can in fact set one of the two values of $\DR$ very large at small lags so to mimic an effective repinning.

\subsection{Realistic drag in the crust}\label{subsec:InterpolDrag}

We now use the methods developed in the previous sections to
construct a non-linear model for superfluid drag in the NS
crust. \citet{Graber2018} have computed the drag coefficient resulting
from Kelvin wave excitations: their estimate of the
drag coefficient depends on the typical relative velocity as
$\tilde\DR\propto |v_{\L\p}|^{-3/2}$. On the other hand,
\citet{Jones1990} has shown that Kelvin processes are suppressed below
$|v_{\L\p}| \lesssim  500$ cm/s, and that at even lower vortex velocities $|v_{Lp}|\approx 1$ cm/s the main contribution to the drag comes from phonon excitations, leading to a constant coefficient $\DR = 10^{-5}$. 

Since phonon excitations give rise to a drag coefficient which is lower than the Kelvin one, a realistic model that interpolates between the two behaviours should reduce to 
\begin{equation}
\begin{split}
    &\tilde\DR \approx\, \DR_1\left(\frac{|\vb_{\L\p}|}{v_1}\right) 
    \quad \qquad \text{for}\, v_{\L\p}\approx v_1 \ll\, 500 \text{cm/s} \\
    &\tilde\DR \approx\, \DR_2\left(\frac{|\vb_{\L\p}|}{v_2}\right)^{-3/2} 
    \quad  \text{for}\, v_{\L\p}\approx v_2 \gg 500\, \text{cm/s} \, ,     
\end{split}
\end{equation}
where $v_1$ and $v_2$ are respectively the typical velocities for which the phonon and kelvin excitations are dominant. 
We choose the simplest linear dependence to get a smooth interpolation between the two channels and the values $\DR_1$ and $\DR_2$  are determined from microphysical calculations of phonon and kelvin drag parameters for $v_{\L\p}\approx v_1$ and $v_{\L\p}\approx v_2$ respectively. We note that \citet{Erbil2020} interpret the result of \cite{Graber2018} differently and that results in a exponent for the Kelvin-drag regime of $-1/2$. Since our model is purely phenomenological, this can be easily adjusted in our prescription.

\begin{figure}
    \centering
    \includegraphics[width = 0.99
    \columnwidth]{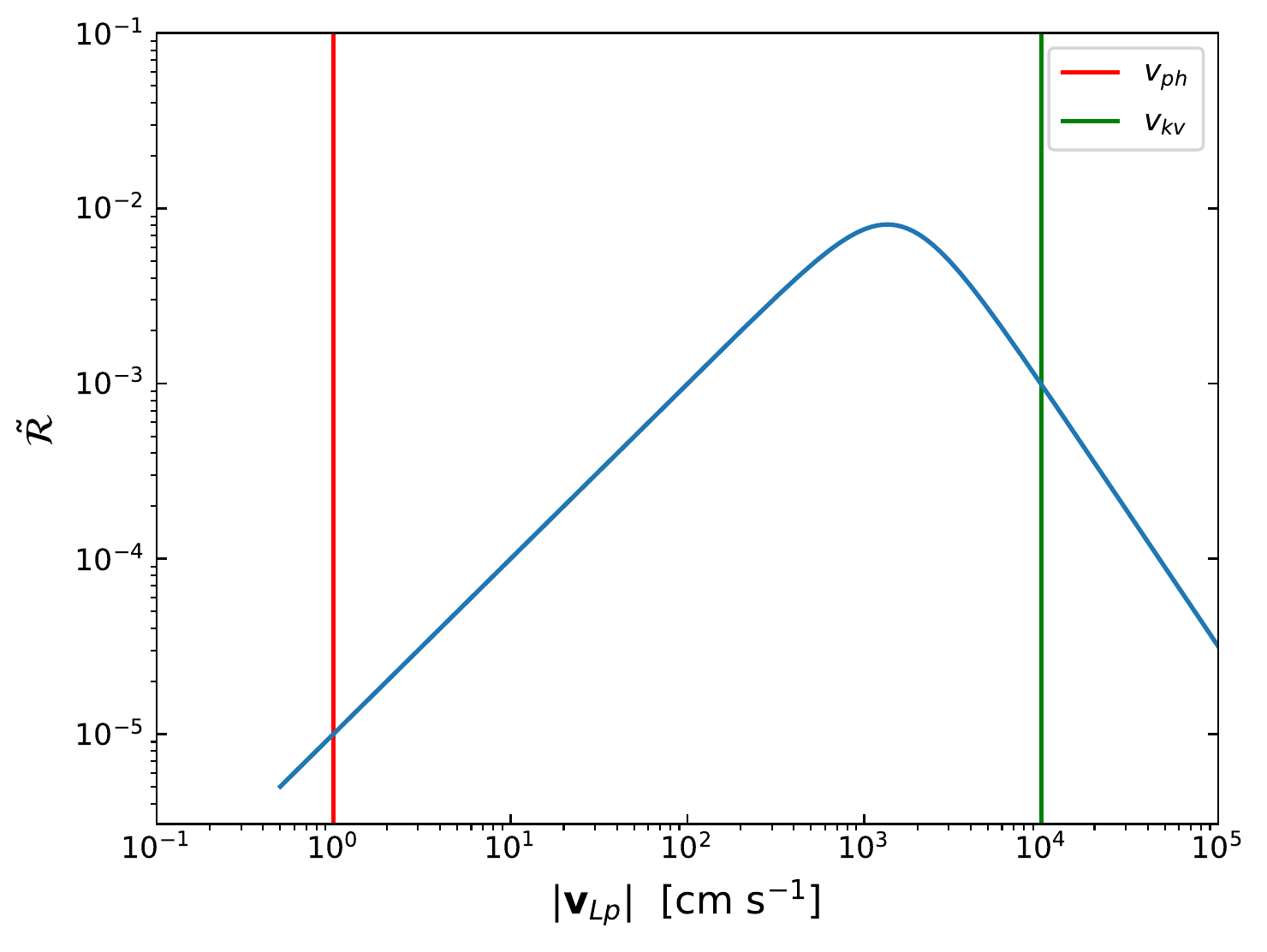}
    \caption{Form of the $\tilde\DR$ as a function of $|\vb_{Lp}|$ where $\DR_1 =10^{-5}$, $\DR_2 = 10^{-3}$, $v_1=1\,$cm/s and $v_2 =10^4\,$cm/s. Since here $\tilde \DR \ll 1$ we have $\mathcal{B}_{MF} \approx \tilde \DR$, so that the figure shows with a good accuracy also the behaviour of the mutual friction parameter.}
    \label{fig:RildeInterpolPicture}
\end{figure}

For our estimate of crustal drag forces we take the constant\footnote{
Clearly, the parameters $\DR_2$ and $v_2$ vary with density, while here we are not doing so because the glitch model will be rigid (the angular velocity will not depend on the $x$ and $z$ coordinates). 
Hence, we choose the value of $\DR_2$ and $v_2$ according to model-A presented in \citet{Graber2018}: model-A is the only one for which the estimated value of $v_2$ lies in the validity regime of kelvonic drag throughout all the crust, namely $v_2>10^2$ cm/s$^{-1}$. According to this model, the value of $v_2$ varies, while $\DR_2$ is almost constant throughout the whole crust.
}
values $v_1=1$ cm/s, with $\DR_1=10^{-5}$ and $v_2= 10^4$ cm/s, for which we set $\DR_2= 10^{-3}$.
Since there are two power law regimes, we model the transition between them\footnote{To avoid unnecessary confusion, let us note that this interpolation is not related to the one introduced at the end of \cref{sec:PowerLawdrag}.} by writing the total drag coefficient as
\begin{equation}\label{eq:reale}
    \tilde\DR = \left( \DR_1^{-1}\,\left(\frac{|\vb_{\L\p}|}{v_1}\right)^{-1} + \DR_2^{-1}\,\left(\frac{|\vb_{\L\p}|}{v_2}\right)^{3/2}\right)^{-1} \, , 
\end{equation}
which is shown as a function of $|\vb_{\L\p}|$ in \cref{fig:RildeInterpolPicture}.
Let us stress that the choice of the crossover velocities $v_1$ and $v_2$ agrees with microphysical estimates \citep{Jones1990,Graber2018}, but remains, nevertheless, rather uncertain. 
However, simulations of interactions between vortices and ions in the crust \citep{Wlazlowski2016} and flux tubes in the core \citep{DrummondI,DrummondII} are  becoming feasible and more stringent constraints on these parameters may become available in the future.

Introducing dimensionless quantities as in the previous subsection,  via \cref{eq:DissipationSine} and taking
\begin{equation}
     \chi_{L}^{(i)}= \frac{|\vb_{\L\p}|}{v_i}\quad , \quad \chi_{np}^{(i)} = \frac{|\vb_{\n\p}|}{v_i} \quad \text{for }i = 1,2  
\end{equation}
where $i$ labels the two crossover velocities, we can write \cref{eq:FormalComponentSolution} as
\begin{equation}\label{eq:split1}
\begin{split}
   &\sqrt{1-s^2} - \left( \DR_1^{-1} \chi_{L}^{(1)\,-1} + \DR_2^{-1} \chi_{L}^{(2)\,3/2} \right)^{-1}\,s = 0 \; ,
   \\
   & s\,\chi_{L}^{(i)} - \chi_{np}^{(i)} + 
   \\
   & \qquad + \left( \DR_1^{-1} \chi_{L}^{(1)\,-1} + \DR_2^{-1} \chi_{L}^{(2)\,3/2} \right)^{-1}\, \sqrt{1-s^2}\,\chi_{L}^{(i)} = 0  \; .
\end{split}
\end{equation}
It is then shown that   $|v_{\L\p}|\leq |v_{\n\p}|$ is still valid, namely $\chi_L^{(i)} = s\chi_{\n\p}^{(i)}$. Hence, it is easy to  recast \cref{eq:split1} as 
\begin{equation}\label{eq:InterpolAlgebraicEq}
    \Bigg( 
    \frac{1}{\DR_1} \left( s\,\chi_{np}^{(1)} \right)^{-1}+\frac{1}{\DR_2}\left(s\,\chi_{np}^{(1)}\,\frac{v_1}{v_2}\right)^{3/2}
    \Bigg)^{-2}\,s^2 + s^2 \, -1 = 0 \,.
\end{equation}
This equation can now be integrated together with the fluid equations for the system, which we will present in the next section.

%%%%%%%%%%%%%%%%%%%%%%%%%%%%%%%%%%%%%%%%%%%%%%%%%%%%
%  %  %  %  %  %  %  %  %  %  %  %  %  %  %  %  %  %  
%%%%%%%%%%%%%%%%%%%%%%%%%%%%%%%%%%%%%%%%%%%%%%%%%%%%

\section{Glitch Model}\label{sec:GlitchModel}

Having determined the model for the mutual friction, we investigate its effect on pulsar glitch dynamics. To do this we start from the full equations of motion for the neutron
superfluid ($\n$) and normal component ($\p$), which can be written  as
\begin{equation}\label{totalEUL}
\begin{split}
	   \big(\partial_t + v_\x^j\nabla_j\big)\big(v_i^\x + \varepsilon_\x\,w_i^{\y\x}\big) &+ \nabla_i\big(\mu_\x + \phi\big) +\\
       &+ \varepsilon_\x\,w_{\y\x}^i\nabla_iv_j^\x = f_i^\x/\rho_\x \, ,
\end{split}
\end{equation}
where $\x$ and $\y$ label the chemical component (i.e. $\x,\y= \n,\p$), $\varepsilon_\x$ 
is the entrainment parameter, $\mu_\x$ is the chemical potential per unit mass of the 
substance $\x$, $\phi$ is the gravitational potential and $w^{\y\x}_i=v^\y_i-v^\x_i$ is the relative
velocity (see e.g. \cite{Prix2004} and \cite{Andersson2006a}). 
To build a rigid glitch model we assume axial symmetry (i.e. no dependence on the azimuthal angle
$\varphi$) and we take the ansatz, with $x$ the cylindrical radius,
\begin{equation}\label{eq:RigidAnsatz}
    \vb_\n \,=\,  x \,\Om_n(t) \, \hat e_\varphi
    \quad , \quad 
    \vb_\p \,=\,  x \,\Om_p(t) \, \hat e_\varphi \, .
\end{equation}
With this assumption the two fluids equations reduce to
\begin{equation}\label{rotating}
\begin{split}
    &\partial_t\,v^\n_i + \varepsilon_\n\partial_t\big(v^\p_i - v^\n_i\big) + \nabla_i\big(\mu_\n + \phi\big) = f^\n_i/\rho_\n \\
     &\partial_t\,v^\p_i + \varepsilon_\p\partial_t\big(v^\n_i - v^\p_i\big) + \nabla_i\big(\mu_\p + \phi\big) = f^\p_i/\rho_\p \,.
\end{split}
\end{equation}
The ansatz \cref{eq:RigidAnsatz} implies that the non-azimuthal components of the two fluid equations in \cref{totalEUL} are not dynamical (they represent the hydrostatic equilibrium that sets the structure of the star) and that the continuity equations for the two species are automatically satisfied \citep{antonelli17}.

The force density that enters the two fluid equations is the mutual friction force, 
\begin{equation}
    f^\n_i /\rho_\n = {F}^{MF}_i/\rho_\n=-\varepsilon_{ijk}\,\omega_\n^j\,\big(v_\L^k - v_\n^k\big)
\end{equation}
Because of the presence of the vorticity $\omega_\n^i$ in the mutual friction, it is sometimes useful to perform a change of variables and define a new angular velocity given by
\begin{equation}
    \label{eq:OmvOmn}
    \Om_v = \big(1 - \varepsilon_\n\big)\,\Om_\n + \varepsilon_\n\,\Om_\p \, .
\end{equation}
This quantity is just a variable related to the superfluid momentum that can be used in place of $\Omega_\n$ and  merely represents the total amount of vortices present within the cylindrical radius $x$ via the Feynman-Onsager relation. 
As a result, the equations of motion in \cref{rotating} read
\begin{equation}
\begin{split}
     &\dot\Om_v  \, = \, - 2\,\Om_v\,\frac{v_L^x}{x} \;, \\
      \Bigg( &\rho_\p - \frac{\varepsilon_\n \, \rho_\n}{1-\varepsilon_\n}  \Bigg) \, \dot\Om_\p  + \frac{\rho_\n}{1-\varepsilon_\n} \, \dot\Om_v
     \, = \, 0  \, ,
\end{split}
\label{equazioniBrutte}
\end{equation}
where we exploited \cref{crespelle}, namely the fact that 
\begin{equation}
    \wn = \frac{1}{m_\n} \nabla \times \pn = 2\Om_v\,\hat e_z \,.
\end{equation}
These are the equations for a two component rigid glitch model, where the
first is a continuity equation for the vortex number  (cfr
eq. (13) of \citealt{antonelli17}) while the second is just the angular
momentum conservation (i.e. it is equivalent to $\rho_\n \dot\Om_\n+\rho_\p \dot\Om_\p=0$). 
More precisely, to obtain a well defined averaged rigid model  we  should average the equations over the whole star, so that the second equation in \cref{equazioniBrutte} 
expresses the conservation of the total angular momentum of the NS,
\begin{equation}
   \dot L_n + \dot L_p = I_n\,\dot\Omega_n + I_p\,\dot \Omega_p = 0\, .
\end{equation}

If we now add an external spin down torque (divided by the total moment of inertia) $\alpha$ due to electromagnetic emission and use \cref{eq:BMF}, the glitch model equations become
\begin{equation}\label{eq:2CompEq}
\begin{split}
        & \dot\Om_v = -2\, \Om_v \, \Bc_{MF}\, \frac{\Om_v -\Om_\p}{1-\varepsilon_\n}
        \\
        & \dot\Om_\p = -\frac{\alpha}{1-x_v} - \frac{x_v}{1-x_v} \dot\Om_v
\end{split}
\end{equation}
where we have introduced the fractional moment of inertia $x_\x = I_\x
/ I_T$ (x = p,n) and $x_v = x_n /(1 - \varepsilon_n)$, while $I_T$ is the total moment of inertia of the star. We also exploited the fact that $\Om_{v\p} = \big(1 - \varepsilon_\n\big)\Om_{\n\p}$. 

Hence, the form of the equations \cref{eq:2CompEq} does not change because of the additional entrainment coupling; moreover, it is possible to include $\varepsilon_n$ into the phenomenological parameters. Since our aim is to study the effect of non-linear mutual friction (and as we have shown that the presence of $\varepsilon_n$ does not change its form), we will set $\varepsilon_n = 0$ so that \cref{eq:OmvOmn} reduces to $\Omega_v = \Omega_n$ and $x_v=x_n$. 

The p-component here represents the `normal' component, i.e. the 
proton-electron fluid in the star, the crust, and all components that
are coupled to it on a dynamical timescale that is shorter than that
of the glitch. In many models it is assumed that, due to electron scattering off magnetised vortex cores \citep{alpar_sauls1988}, the superfluid in the
core is coupled to the crust fast enough that it can be included in
the $\p$ component. However, in the outer core the coupling timescale due to mutual friction may be comparable with the rise time \citep{newton_etal15}. 
This effect has been studied both by integrating the full density dependent equations with also density dependent mutual friction in the core and constant drag in the crust \citep{HP12,HaskellDanai2013} or by treating the superfluid in the core as an additional component in a three component model with density dependent drag in the crust \citep{Graber2018,pierre3comp,SourieChamel2020}. 
What is observed is that the outer core recouples after the glitch, giving rise to a short term relaxation and possibly an `overshoot', in which the observed frequency rises above the observed long term post-glitch frequency, a behaviour that has indeed been observed in a recent glitch of the Vela pulsar \citep{ashton2019Nat,pierre3comp}. 

In the following we adopt an approximate prescription to model this behaviour by modifying equations  \cref{eq:2CompEq} to account for a third component - with fractional moment of inertia $r x_\p$ - that recouples to the remaining part of the  p-component (having fractional moment of inertia $(1-r)x_\p$) with a typical timescale $\tau_{co}$. We thus consider the following system of equations
\begin{equation}
\label{eq:3CompEq}
\begin{split}
        & \dot\Om_\n = - 2 \Bc_{MF} \Om_\n \, ( \Om_\n -\Om_\p  ) \;,
        \\
        & x_\p \big(1 - r\,e^{-t/\tau_{co}}\big) \,\dot\Om_\p + x_\n \,\dot\Om_\n
        = -\alpha  \; ,
\end{split}
\end{equation}
which imply that the third component is completely decoupled at $t=0$: 
at the beginning of our simulation the effective moment of inertia fraction of the normal
component is $x_\p(1-r)$, lower than the asymptotic value $x_\p$ that is reached at the 
end of the recoupling. Clearly, the general relation $x_\p + x_\n =1$ is still valid.

Following \cite{HaskellDanai2013} and \cite{WGNewton2015} we will set
$r=0.75$ and choose for $\tau_{co}$ the fiducial value of $71$ sec so that the latter is compatible with the standard value of the mutual friction coefficient in the core. Later on we will test how a different choice of $\tau_{co}$ affects the glitch sizes predicted with this model.

%%%%%%%%%%%%%%%%%%%%%%%%%%%%%%%%%%%%%%%%%%%%%%%%%%%%
%  %  %  %  %  %  %  %  %  %  %  %  %  %  %  %  %  %  
%%%%%%%%%%%%%%%%%%%%%%%%%%%%%%%%%%%%%%%%%%%%%%%%%%%%

\section{Numerical results: study of the glitch rise}

\begin{figure}
    \centering
    \includegraphics[width = 0.99 \columnwidth]{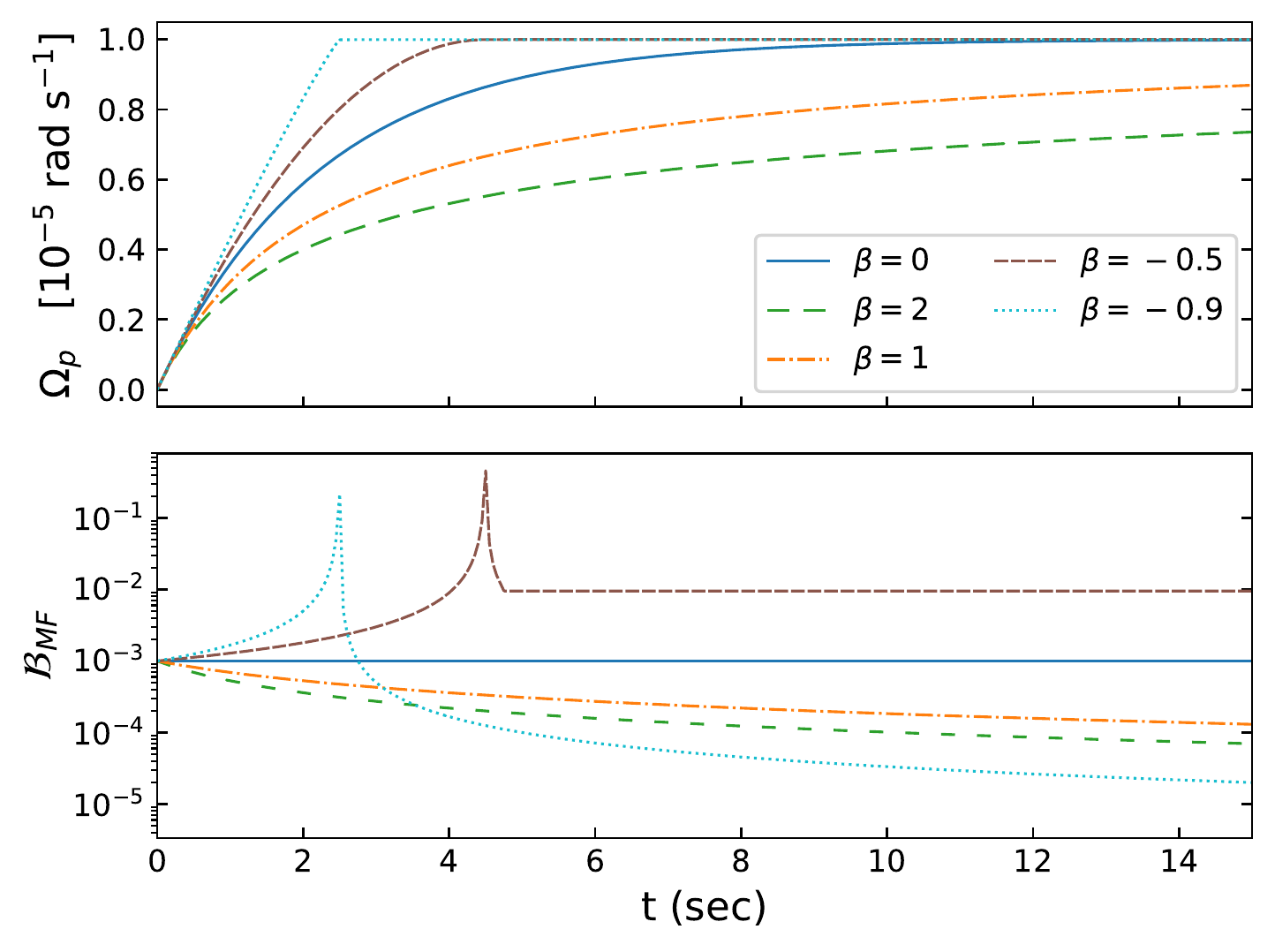}
    \caption{
    Study of the rise time for a power law drag $\tilde\DR$ with initial lag 
    $\Om_{\n\p} = 10^{-3}\,$rad/s , $x_\p = 0.99$ and torque $\alpha = 10^{-9}\,$rad/s$^2$. 
    }
    \label{fig:BetaStudy}
\end{figure}

To begin our analysis we perform a numerical integration of  the two-component model defined by the equations in \cref{eq:2CompEq} for the  power law ansatz \cref{eq:PowerLawAlgebraicEq}. 
First, we study how the glitch rise time changes with the value of $\beta$: the results are shown in \cref{fig:BetaStudy}. 
Following \citet{seveso_etal16}, we assume that initially there is a lag between the two fluids of $\Om_{\n\p} = 10^{-3}\,$rad/s, which, for a stellar radius of $10\,$km, corresponds to a lag of $v_{\n\p} = 10^3\,$cm/s in the crust near the equator. 

To compare cases corresponding to different values of $\beta$, we impose that all models have the same value of $\tilde\DR=10^{-3}$ at $t=0$, so that the initial slope of the rise is equal for each value of $\beta$ because, initially, the angular momentum is transferred with the same $\Bc_{MF}$. 
We then simulate each model by solving the implicit equation \cref{eq:PowerLawAlgebraicEq} at each integration step so that the evolution of $\tilde\DR$ and $\Bc_{MF}$ is peculiar to each model. 
With the simple power law model we cannot test values of $\beta \le -1$ (and thus the value $\beta = -3/2$ associated with kelvin waves), because the drag force \cref{eq:manibucate} would diverge when the lag goes to zero, and consequently \cref{eq:PowerLawAlgebraicEq} may not have solutions. 
This issue is addressed in the more realistic model we will discuss later in this section.

\begin{figure}
    \centering
    \includegraphics[width = 0.99 \columnwidth]{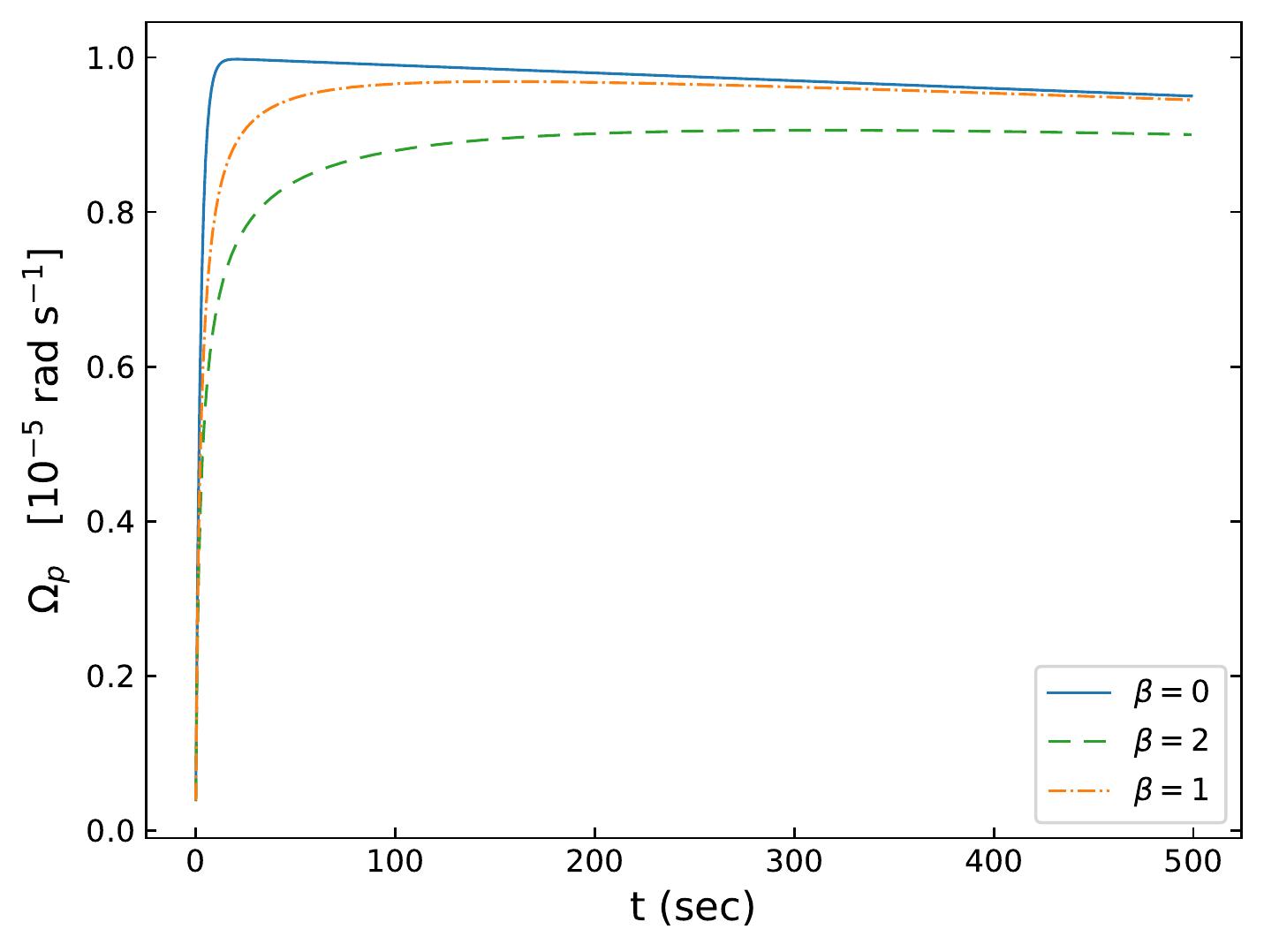}
    \caption{
    Long time evolution for positive $\beta$, in which we verify that the value of the glitch amplitude after the first minutes is not necessarily the same for all cases. 
    The initial lag used is $\Omega_{\n\p} = 10^{-3}\,$rad/s, $I_p / I_{T} = 0.99$ and torque $\alpha = 10^{-9}\,$rad/s$^2$.  }
    \label{fig:VeryLongTime}
\end{figure}

We observe that for negative values of $\beta$ the rise is faster than for the usual linear model with $\beta=0$, which is used here as a reference since it allows to define the exponential timescale for the process. 
Furthermore, $\Om_\p$  grows to the asymptotic value in a finite time because of the rise in the mutual friction coefficient $\mathcal{B}_{MF}$. 
Afterwards the value of the $\mathcal{B}_{MF}$ drops sharply (as can be seen in the lower panel of \cref{fig:BetaStudy}) and the frequency evolution essentially stops. 
Conversely, for positive values of $\beta$ the rise is much gentler because the angular momentum transfer rate decreases with the lag; we also observe that, for values of $\beta$ high enough, this may result in smaller glitches when the time taken for the rise becomes longer than the spin down typical timescale. This effect can be seen in \cref{fig:VeryLongTime} for the $\beta = 2 $ case, where the integration is performed long enough that the spin-down torque effects become apparent.

The behaviour for negative values of $\beta$, for which we observe a very rapid rise, is consistent with observations of recent glitches in the Crab pulsar, and also with a recent large glitch in the Vela pulsar, for which an upper limit on the rise time of $12\,$s was set at the 90\% confidence level, with the data favouring, in general, very short rise times \citep{ashton2019Nat}. 
It is therefore clear that to obtain quantitative constraints on interior NS physics it is not sufficient to calculate drag coefficients for a fixed $|\vb_{Lp}|$ and then treat them as constants to obtain an exponential rise, as the behaviour inferred from timing observations is qualitatively different. At the same time, we have to point out that the current data do not allow to resolve the glitch behaviour below $10\,s$ where the differences with the exponential rise are most visible. This situation might improve in the future, as the observational time span keeps growing- thus allowing for more refined statistical models of the intrinsic noise and improving the sensitivity.

We now turn our attention to the more microphysical prescription for the drag given 
in \cref{eq:reale} and plotted in \cref{fig:RildeInterpolPicture}, 
which allows for both kelvin and phonon contributions in the NS crust. 

From our previous analysis of the power law case, we expect to have
different behaviours for the glitch rise if the initial lag
is larger or smaller than the value for which the maximum of $\Bc_{MF}$ occurs, i.e. if the drag is in
the negative $\beta$ regime for high values of $|\vb_{\L\p}|$, or in the
positive $\beta$ regime for low values of $|\vb_{\L\p}|$. 

\begin{figure}
    \centering
    \includegraphics[width = 0.99 \columnwidth]{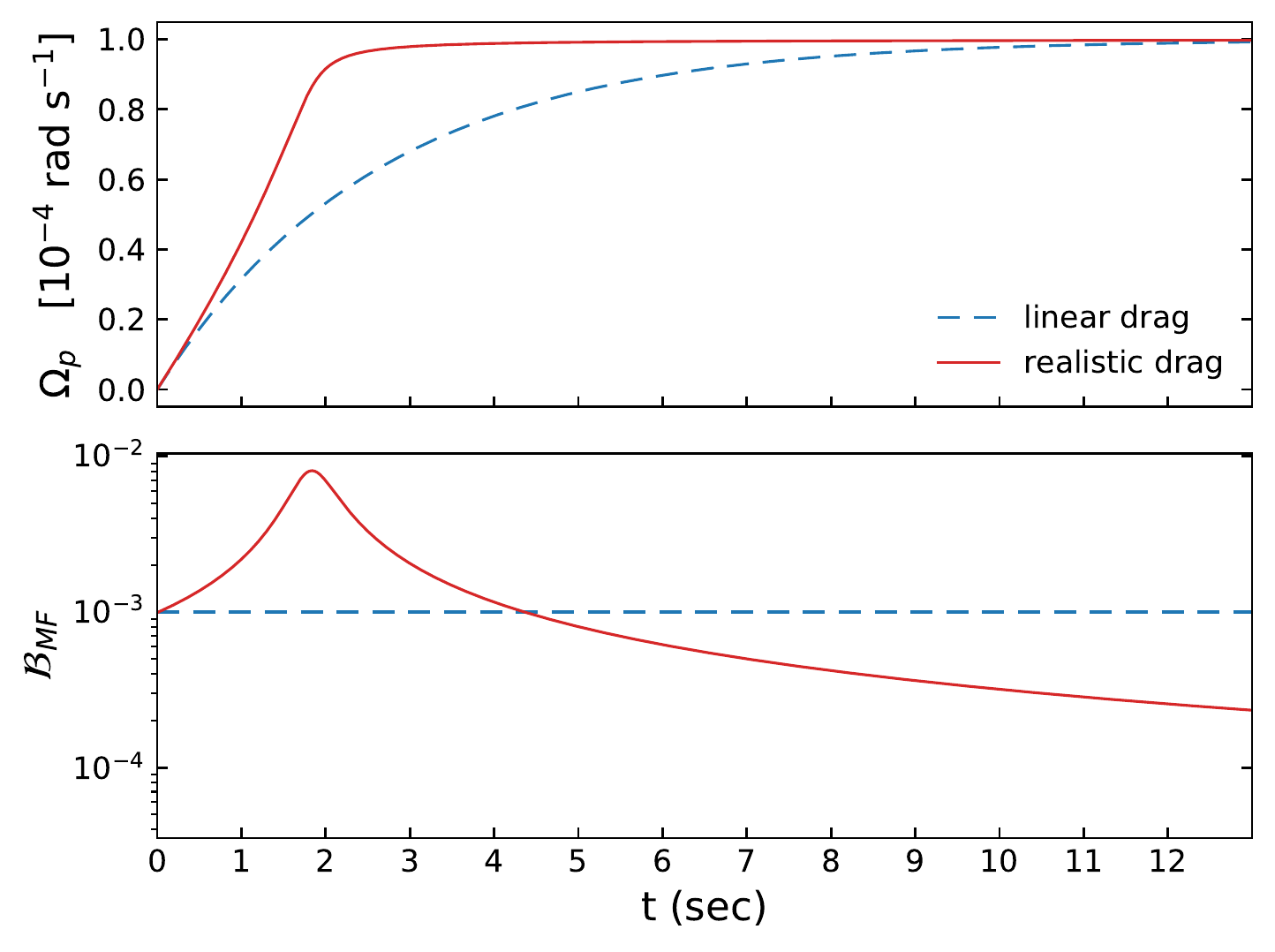}
    \caption{    Glitch rise (red, solid) for the microscopic drag prescription in \cref{eq:reale}  compared to the exponential rise given by linear mutual friction (blue, dashed) for $x_p=0.99$. The initial lag is $\Delta\Omega_{np} = 10^{-2}\,$rad/s, larger than the value for which the maximum of $\Bc_{MF}$ occurs.}
    \label{fig:InterpolBeforePeak}
\end{figure}

\begin{figure}
    \centering
    \includegraphics[width = 0.99 \columnwidth]{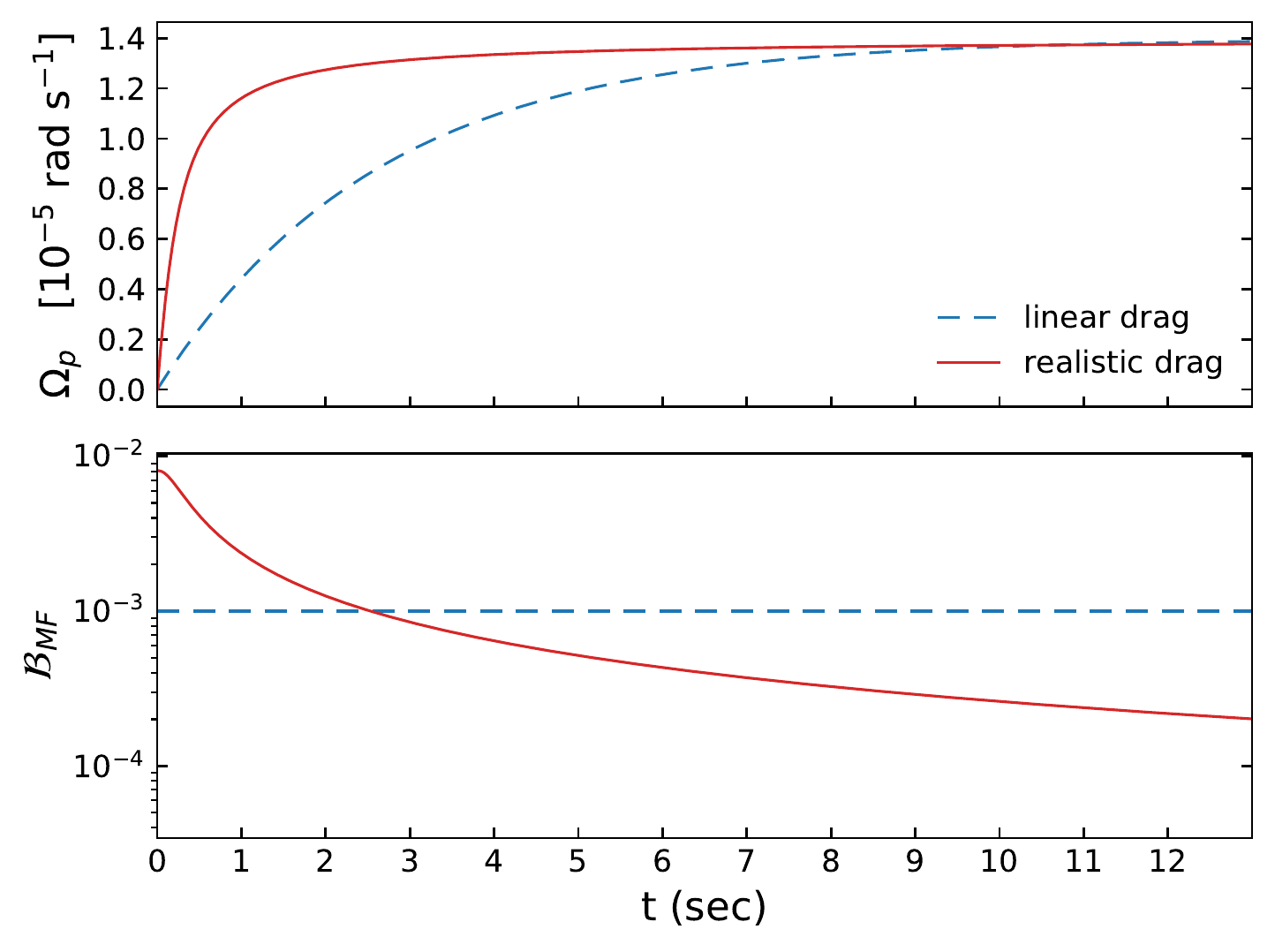}
    \caption{
    Glitch rise (red, solid) for the microscopic drag prescription in \cref{eq:reale}  compared to the exponential rise given by linear mutual friction (blue, dashed) for $x_p=0.99$. 
    The initial lag is $\Om_{\n\p} = 1.4 \times 10^{-3}\,$rad/s that corresponds to the value for which the maximum of $\Bc_{MF}$ occurs.}
    \label{fig:InterpolTopPeak}
\end{figure}

\begin{figure}
    \centering
    \includegraphics[width = 0.99 \columnwidth]{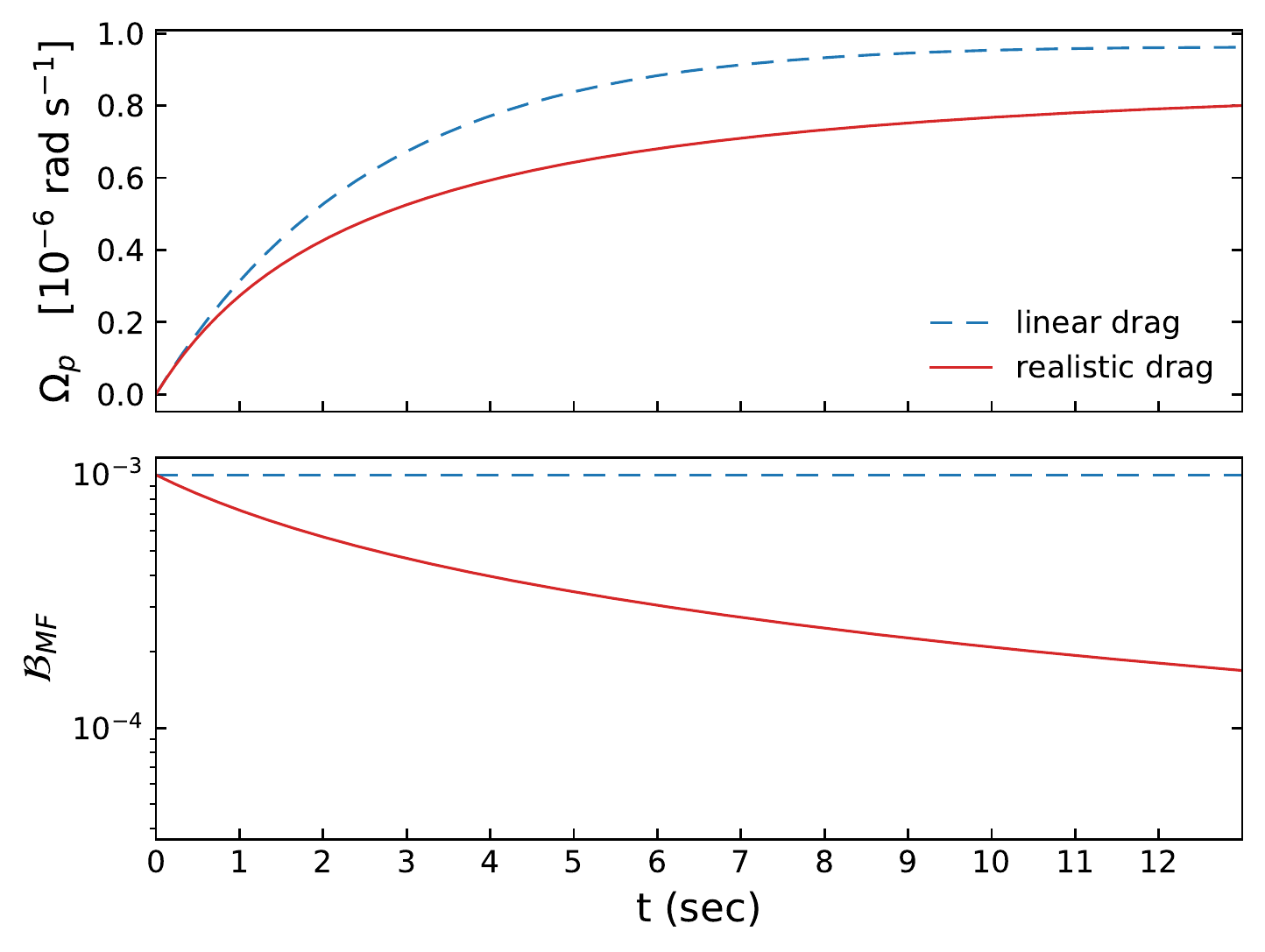}
    \caption{
    Glitch rise (red, solid) for the microscopic drag prescription in \cref{eq:reale} compared to the exponential rise given by linear mutual friction (blue, dashed) for $x_p=0.99$. The initial lag is
    $\Om_{\n\p} = 10^{-4}\,$rad/s and thus smaller than the value for which the maximum of $\Bc_{MF}$ occurs.}
    \label{fig:InterpolAfterPeak}
\end{figure}

To investigate this we integrate the two-component model for different initial
conditions, for initial values of $v_{np} = R \Omega_{\n\p}$ both before and after the peak of $\Bc_{MF}$. 
Examples of the results are shown in \cref{fig:InterpolBeforePeak,fig:InterpolTopPeak,fig:InterpolAfterPeak}.
In the case with a large lag at $t=0$ that falls in the $\beta<0$ part of the drag, we can observe a change in the convexity of the rise in correspondence of the activation of the phonon dissipation channel, see \cref{fig:InterpolBeforePeak}. 
In these cases the initial rise is very rapid, and likely to decouple part of the core, causing an `overshoot' and rapid post-glitch recoupling, as expected in some glitch models \cite[see e.g. ][]{HP12,antonelli17,Graber2018,pierre3comp}. 
For small values of the initial lag, on the other hand, only values of $\beta>0$ are sampled, and the mutual friction strength drops off as the glitch proceeds. 
Like in the previous simpler case, this behaviour can possibly lead to smaller amplitude glitches, as the coupling timescale becomes long enough to be comparable with the spin-down timescale and the rise is effectively halted.

Throughout this entire section (and in the following one as well) we used $x_\n=0.01$. This value of the fractional moment of inertia is consistent with that of a crustal superfluid, even though the S-wave superfluid might well extend into the inner core \citep{Zuo2004} - in which case a value of $x_\n \approx 0.1$ would be more appropriate. Still, we decided to use $x_\n = 0.01$ to be consistent with the realistic model developed for the crust. A larger value of the superfluid fractional moment of inertia would result in a larger angular momentum reservoir, and therefore bigger glitches.

\section{Numerical Results: Glitch size distributions}

\begin{figure}
    \centering
    \includegraphics[width = 0.99 \columnwidth]{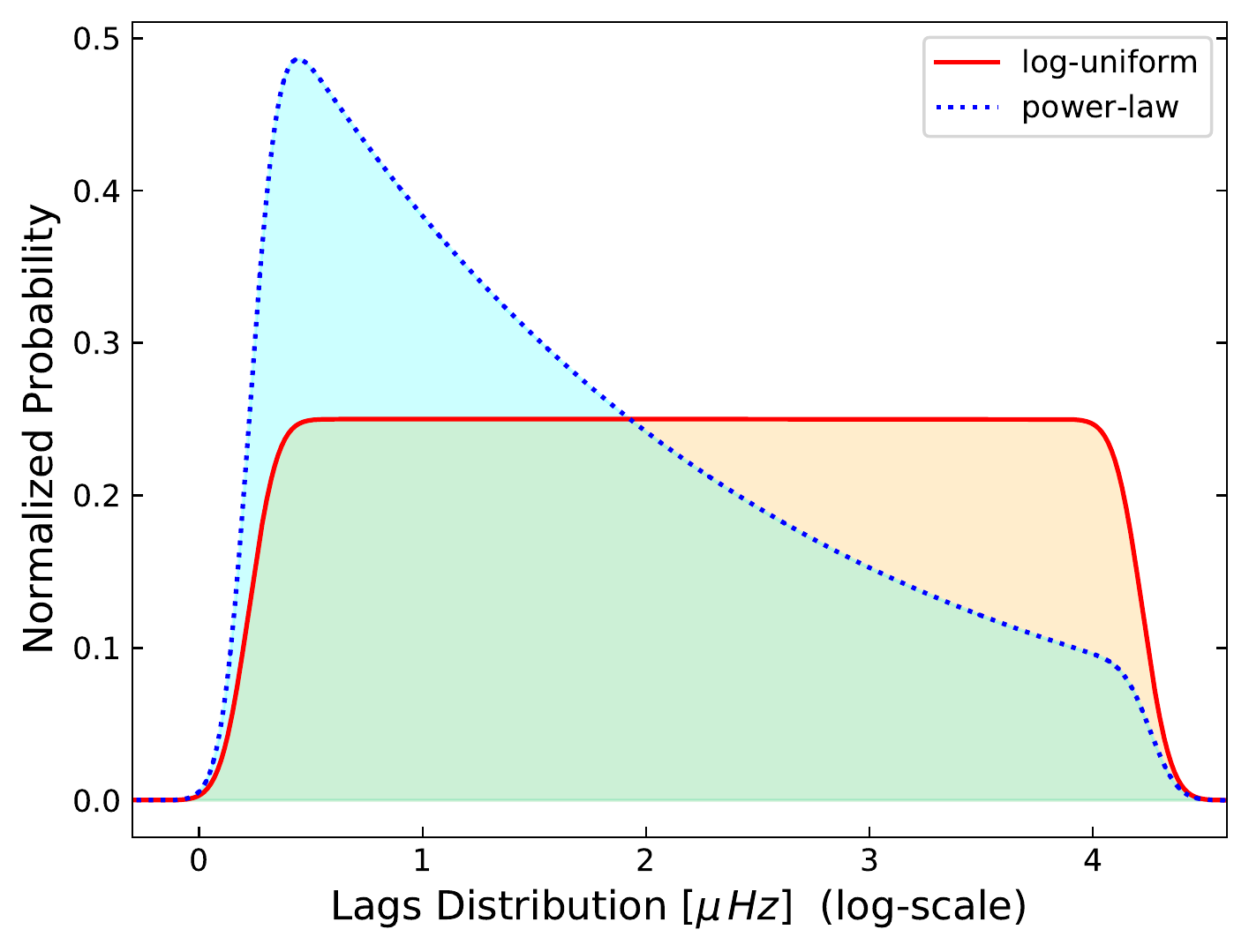}
    \caption{Input distributions of the initial lags for the two cases of interest: log-uniform (red, solid) in the interval $[10:10^5]$ cm/s and power law (blue, dashed) in the same interval with exponent $-1.2$. }
    \label{fig:LagsDistrib}
\end{figure}

In the previous section we mostly focused on the effects of the microscopic drag in \cref{eq:reale} on the glitch rise, namely on the first few seconds after the glitch is started. 
The same  friction model can also be used to study the long time frequency evolution of the star after a glitch,  and also glitch sizes. 
The frequency evolution is ultimately determined once the initial lag, and consequently the initial coupling timescale between the two components, is chosen. 
This quantity is not constrained, and may vary from glitch to glitch even in the same star. 
Therefore, we now study the glitch size distributions that we get out of the model for an initial lag in the interval [$10:10^5$] cm/s, see \cref{fig:LagsDistrib}.

First, we observe that for large initial lags the drag is mainly due to Kelvin excitations and increases as the system recouples, leading to a very rapid rise. This in turn may decouple part of the core and lead to a larger initial jump in frequency.
To investigate this effect we consider the approximate 3-component model in \cref{subsec:InterpolDrag}, in which the core recouples with a timescale $\tau_{co}$. 
This means that the angular momentum is transferred to an observable p-component with smaller inertia and results in larger glitches (although note that it will not reproduce an overshoot as a full
3-component model such as that of \citet{Graber2018,pierre3comp,Sourie3Comp}. 
The effect is more evident for large initial lags for  which angular momentum is transferred more rapidly ($\Bc_{MF}$ higher) and the fraction of core that has already recoupled is smaller.  

The initial condition have been chosen to be consistent with parameters of the Vela pulsar, so that $\Omega_p(0) \approx \Omega_n(0) = 70.34$ rad/s and $\alpha = 9.8\cdot10^{-11}$ rad/s$^2$. Also, in order to test the effects of the of $\tau_{co}$ we consider the fiducial value for Vela \citep[see ][]{HaskellDanai2013, WGNewton2015}, namely $\tau_{co}=71$ s, but also $\tau_{co}=7.1$ s which is compatible with the linear model timescale and $\tau_{co}=710$ s. 
Throughout this entire section the glitch is computed from the residuals, namely as $\Omega_p(t) - \Omega_p(0) + \alpha\, t$, where $\alpha$ is the absolute value of the spin down rate, because this is compatible with the observational procedure.

\begin{figure}
    \centering
    \includegraphics[width = 0.99 \columnwidth]{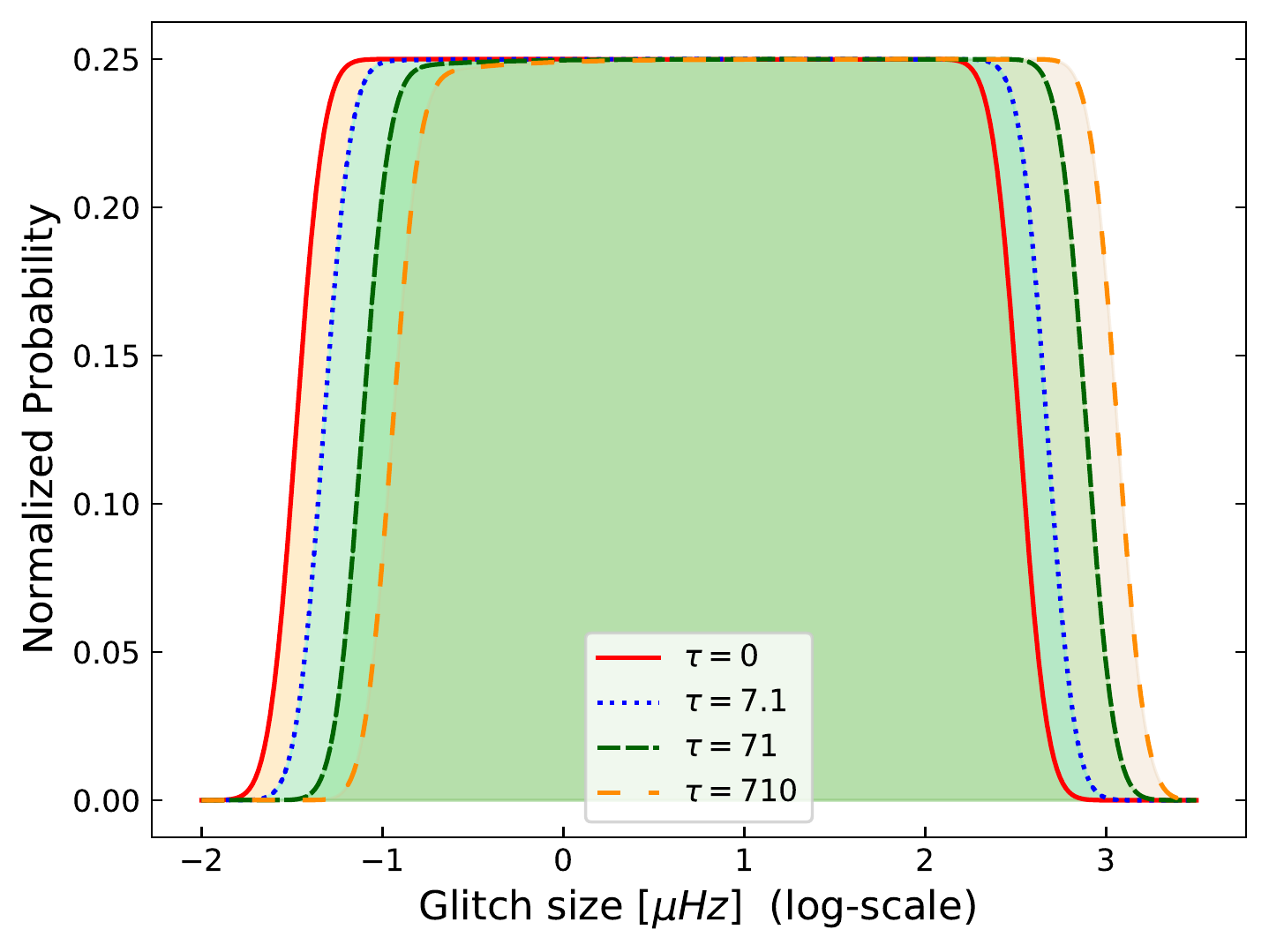}
    \caption{
    Output glitch size distribution in the linear case for an initial (log-scale) uniform lags distribution. The linear $\tau=0$ case (red, solid) is given for comparison with the approximate 3-component ones: $\tau=7.1$ (blue, dotted), $\tau=71$ (green, dashed) and $\tau=710$ (yellow, dashed).}
    \label{fig:VelaFlatLin200}
\end{figure}

As a first step we consider initially a log-uniform distribution of the initial lags $v_{\n\p}$ in the range $[10;10^5]$ cm/s, see \cref{fig:LagsDistrib}. The output glitch distributions in the linear case are presented in \cref{fig:VelaFlatLin200}: as expected they are flat as well and roughly scaled by a factor $x_n = 0.01$, while the only effect of the approximate three component model is to slightly shift the glitch sizes to bigger values. 
Furthermore, since in the linear model the angular momentum is transferred at a constant rate, the effects of the approximate 3-component model are the same for each initial lag, namely the output glitch-size distribution is ``rigidly'' shifted to the right by a greater amount for larger values of $\tau_{co}$. 

\begin{figure}
    \centering
    \includegraphics[width = 0.99 \columnwidth]{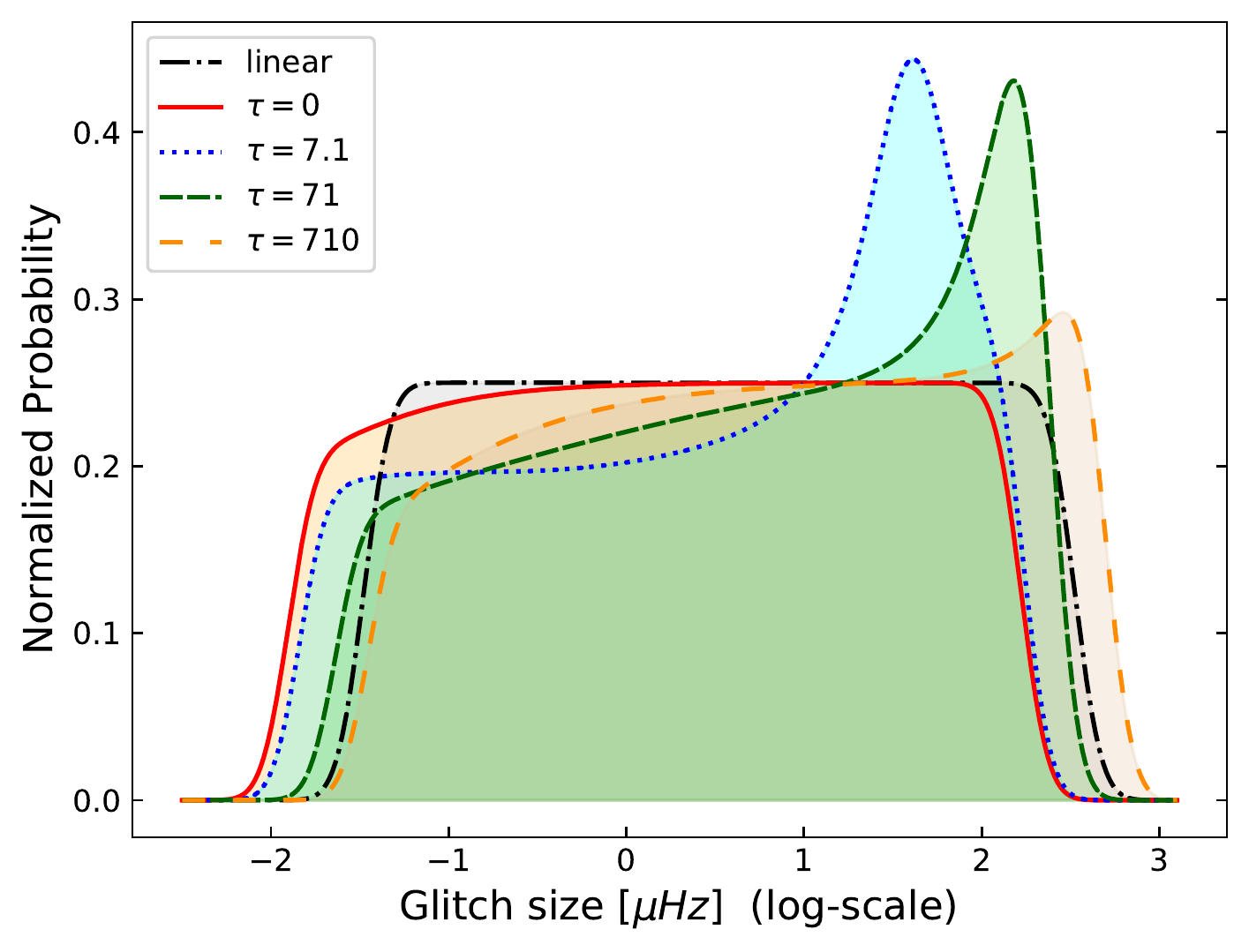}
    \caption{Comparison of glitch size distributions in the linear and realistic non-linear case for an initial log-uniform lags distribution.The linear case (black, dash-dotted) is given for comparison with the non-linear 2-component model (red, solid) and  approximate 3-component ones: $\tau=7.1$s (blue, dotted), $\tau=71$s (green, dashed) and $\tau_{co} = 710$s (yellow, dashed). The glitch is measured at 200 sec.}
    \label{fig:VelaFlatNL200}
\end{figure}

In \cref{fig:VelaFlatNL200} we compare the output glitch distribution for the standard linear drag 2-component model with the ones obtained using the realistic non-linear drag from \cref{subsec:InterpolDrag} and assuming to ``measure'' the glitch after $200$ s. 

\begin{figure}
    \centering
    \includegraphics[width = 0.99 \columnwidth]{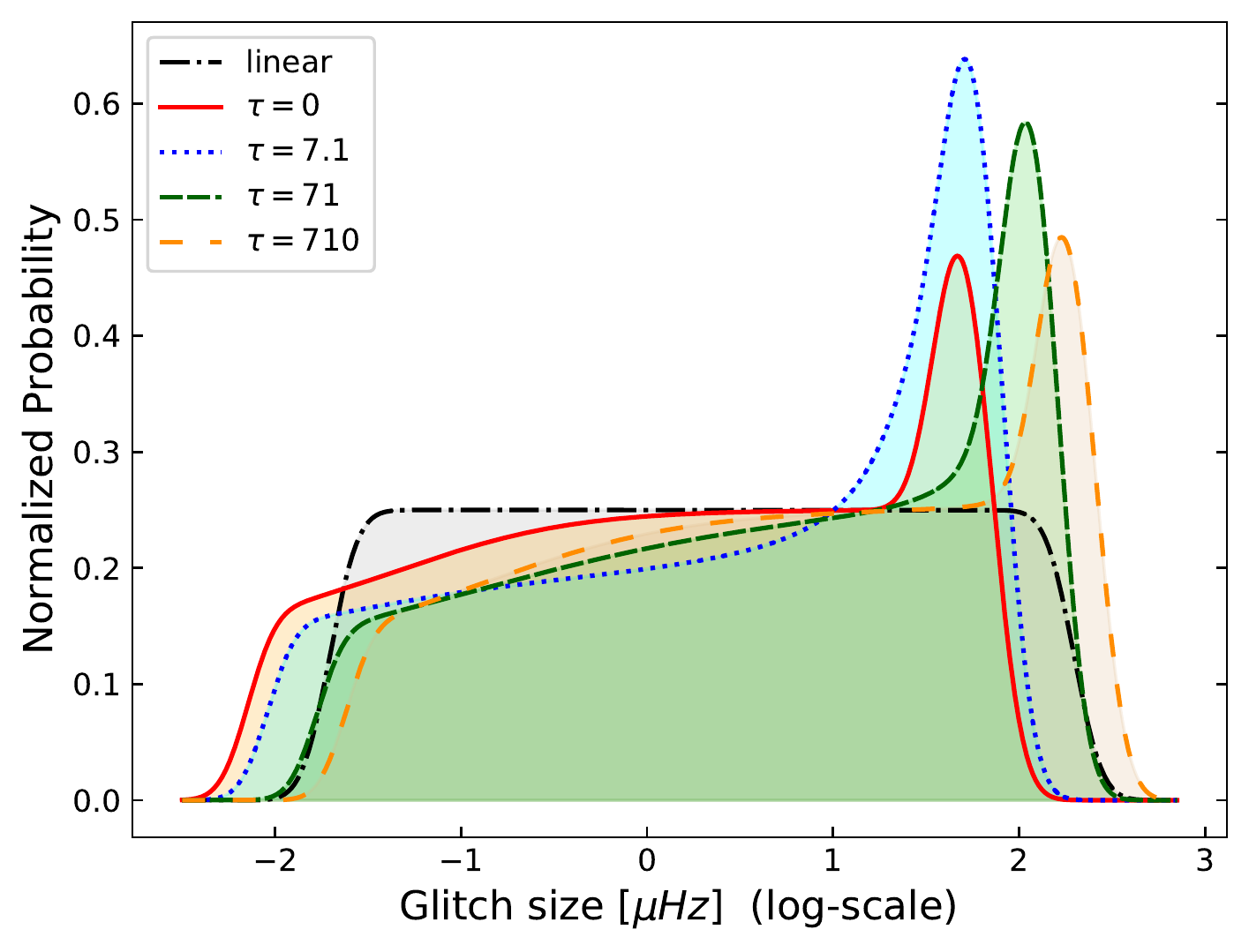}
    \caption{Comparison of glitch size distributions in the linear and realistic non-linear case for an initial log-uniform lags distribution. The linear case (black, dash-dotted) is given for comparison with the non-linear 2-component model (red, solid) and  approximate 3-component ones: $\tau=7.1$s (blue, dotted), $\tau=71$s (green, dashed) and $\tau_{co} = 710$s (yellow, dashed). The glitch is measured at 50 sec.}
    \label{fig:VelaFlatNL50}
\end{figure}

We see that now in all cases with non zero $\tau_{co}$ the output distributions present a peak for high glitches, and while there is a quantitative difference between the different cases, the qualitative feature is present in all the three models.
To point out that this effect is due to the modified mutual friction - and not to the approximate 3-component model, so that it must be visible even for $\tau_{co}=0$ - in \cref{fig:VelaFlatNL50} we plot the output distribution that we get if we ``measure'' the glitch after $50$ sec. Note that the latter is compatible with the current observational limits on the full glitch rise time \citep{dodson2002}, although the initial rise may be of the order of 12 s \citep{ashton2019Nat}. 
In \cref{fig:VelaFlatNL50} the peak is in fact present also for the non-linear $\tau_{co} = 0$ sec model.
This feature is interpreted as follows: for very high initial lags, namely $v_{np}\approx10^5\,$cm/s, the system is sampling the area well right to the peak of the $\Bc_{MF}$ plot (see \cref{fig:RildeInterpolPicture}), so that we have a considerably low value for the initial angular momentum transfer rate. This implies that the angular momentum reservoir is not completely emptied out within $50$ s and we measure a smaller glitch size. The system is sampling the kelvonic branch, so that the rate increases as the system recouples and the residual is transferred within the next $150$ s in the $\tau_{co} = 0 $ case - and can possibly be considered as a delayed rise\footnote{
    The delayed rise in the largest glitch observed in Crab pulsar \citep[see][]{shaw2018crab} could be the effect of a non-linear mutual friction -  with model parameters different from those considered in the present work.}. 
If we consider the approximate 3-component model, the dynamics depends on the interplay between two ``timescales'', the angular momentum transfer rate (which changes during the evolution) and $\tau_{co}$. 
As a result, we still observe the peak after 200 s. 
As expected, the microscopic drag of \cref{eq:reale} gives also smaller glitches (with respect to the linear case) when the system samples the phonon-branch only, namely for low initial lags\footnote{
    Recall that $v_{Lp} \le v_{np}$, so that for small initial lag the system is sampling the region left to the $\mathcal{B}_{MF}$ peak, see \cref{fig:RildeInterpolPicture}.}. 
This effect is counteracted in the approximate three component model and in the fiducial model (the one with $\tau_{co}=71$ s) we observe a reduction in the number of small glitches, see \cref{fig:VelaFlatNL200}.

\begin{figure}
    \centering
    \includegraphics[width = 0.99 \columnwidth]{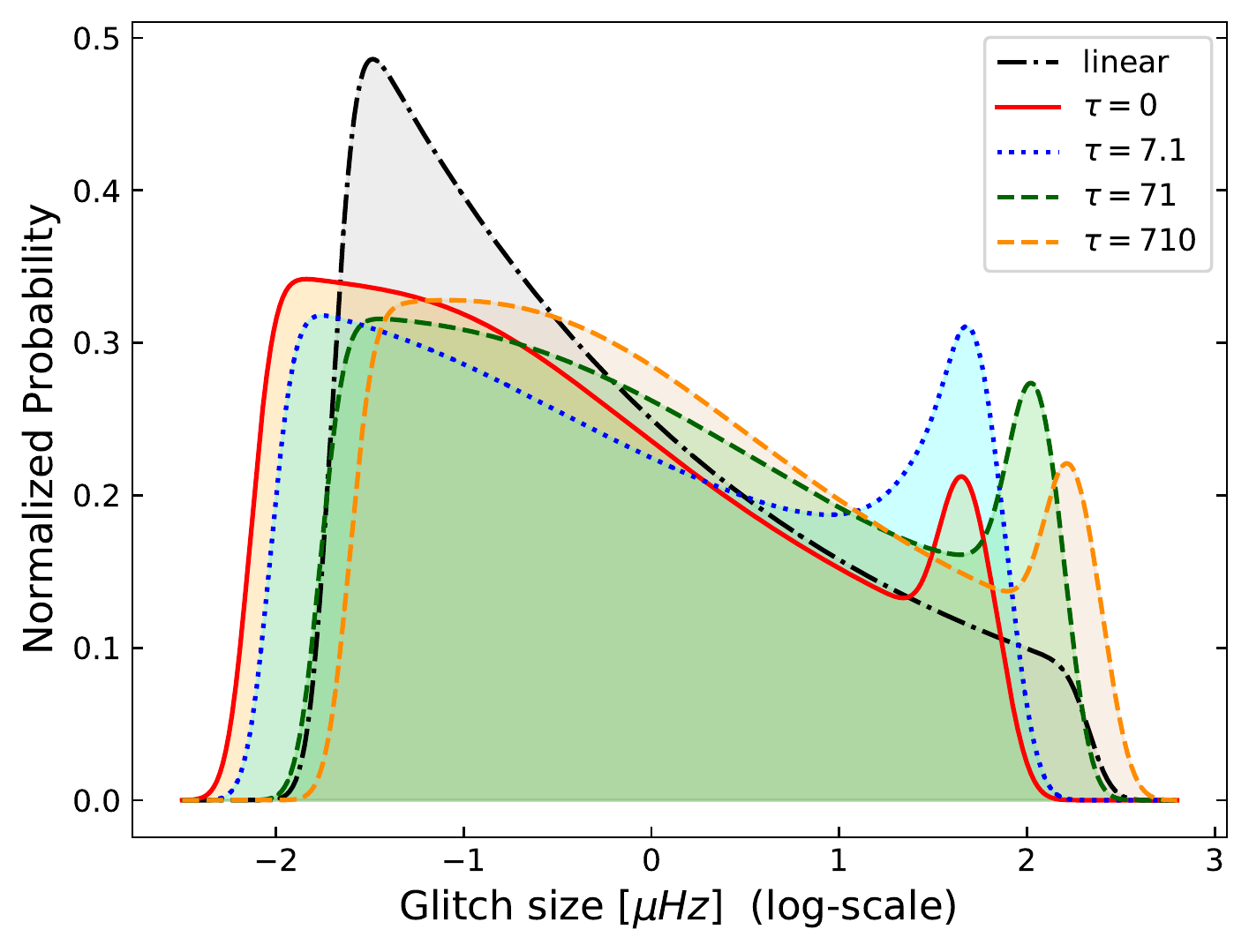}
    \caption{Comparison of glitch size distributions in the linear and realistic non-linear case for an initial power law lags distribution. The linear case (black, dash-dotted) is given for comparison with the non-linear  2-component mode (red, solid) and two  approximate 3-component ones: $\tau=7.1$s (blue, dotted), $\tau=71$s (green, dashed) and $\tau_{co}=710$s (yellow, dashed). The power law exponent is -1.2 and the glitch is measured at 50 sec.}
    \label{fig:VelaPLNL50}
\end{figure}

In \cref{fig:VelaPLNL50} we test a somewhat more realistic scenario in which the initial lag distribution is not log-uniform but follows a power law with exponent $-1.2$ as suggested by both  simulations of vortex avalanches \citep{Warszawski2012a} and observations of pulsar glitch size distributions \citep{HowittApJ2018, MelatosApJ2008, Fuentes2019}, see \cref{fig:LagsDistrib}. 
The features discussed in the previous log-uniform case are still present, but now we are sampling smaller values for the initial lags with higher probability.  
In this case the distribution turns out to be bimodal, with a narrower larger size component above approximately  $10^{2} \,\mu Hz$ and a wider second component extending to lower sizes. 
This distribution is in qualitative agreement with the observed distribution presented by \citet{fuentes17} and \citet{ashton17}, which suggests that the same mechanism (the recoupling of a pinned superfluid) can explain both populations if a non-linear drag model is used. 

It is thus possible that in NSs with predominantly large glitches (like the Vela pulsar) the events are triggered in stronger pinning regions, while in stars with a majority of smaller glitches, these are likely to be triggered in weaker pinning regions. 
This is in agreement with the analysis of \citet{haskell2018corecrust}, according to which larger glitches are likely to be triggered in the outer core of the star. To implement this feature in the model one could make use of a distribution for $x_\n$ (instead of a flat fixed value $x_\n = 0.01$). In this way one can account for the fact that, for large values of the lags, the superfluid reservoir involved in the process extends also to the outer core -were it is pinned- and therefore its fractional moment of inertia is bigger. Given the uncertainties in the values for $x_\n$ we have not included this feature in the model, but this constitutes an interesting refinement to be explored in future works.

Although we have also tested approximate 3-component models, none of them will reproduce an important feature of the full 3-component ones - the overshoot \citep[see][]{Graber2018,Sourie3Comp,pierre3comp}. The presence of an overshoot can lead to a glitch size (measured at $50\,s$) that is bigger than the asymptotic value. The overshoot is expected to affect more for large initial lags, and therefore it would shift the rightmost peak by some amount. However, according to up-to-date estimates \citep[see][]{Ale2016Bayes} the overshoot size will not be bigger then double the asymptotic value, that is it will shift the rightmost peak by no more than $\approx 0.3$, so that the qualitative features are the same. To this respect, we also note that the asymptotic value of our approximate 3-component model is bigger than the 2-component one, and in some way already accounts for the overshoot effects in the size distributions by slightly overestimating them (see \cref{fig:VelaPLNL50}, especially the fiducial $\tau = 71$ case).

Also, to obtain more stringent constraints future work should include corrections due to general relativity \citep[see e.g. ][]{SC16,antonelli17b,gavassino2020universal} and the full density dependence of the drag in the core \citep{alpar84rapid,Andersson2006a} and crust of the star \citet{Graber2018}.
%

%%%%%%%%%%%%%%%%%%%%%%%%%%%%%%%%%%%%%%%%%%%%%%%%%%%%
%  %  %  %  %  %  %  %  %  %  %  %  %  %  %  %  %  %  
%%%%%%%%%%%%%%%%%%%%%%%%%%%%%%%%%%%%%%%%%%%%%%%%%%%%

\section{Conclusions}

We studied the effect of a non-linear form of the mutual friction on pulsar glitch sizes and rise times: we  considered both a simple power law dependence of the drag force on the relative velocity between superfluid vortices and the normal component, and a physically motivated model for the crust of a NS, in which for low vortex velocities the drag is mainly due to phonon excitations while for high velocities to kelvin excitations \citep{Jones1990,Jones1992,epstein_baym92,Graber2018}.

For the simple power law case we find that for positive values of the index $\beta$ the rise is slower than in the standard linear case, for which the rise is exponential. 
This case is relevant for both classical turbulence, for which one expects a power law index $\beta=1$ and isotropic quantum turbulence for which $\beta=2$ and confirms the previous results of \cite{peralta2006} for the rise time. Additionally we find that for $\beta>0$ the size of the glitch can also be affected, as the coupling timescale rapidly grows to the point where it is comparable with the spindown timescale, thus effectively halting the rise and leading to smaller glitches.

For negative values of the power law index $\beta$  the situation is reversed, and the drag parameter $\tilde{\mathcal{R}}$ grows as the vortices slowdown. 
This is particularly relevant for glitches, as the rapid rise is thought to be due to mutual friction coupling given by the excitation of Kelvin waves either in the crust, as vortices move past the ions in the lattice, or core, as they cut through superconducting fluxtubes \citep{Ruderman1998,Link2003L}. In both the cases where the dissipation is due to excitation of Kelvin waves the index is expected to be $\beta=-3/2$. 

In the  more realistic model defined in \cref{eq:reale} we have that at high velocities (above $v_0\approx 10^3$ cm/s) vortices experience a kelvonic drag that scales as $v_{\L\p}^{-3/2}$, while at lower velocities the kelvonic contributions are suppressed and phonons dominate the drag, scaling as $v_{\L\p}$ \citep{Jones1990}. 
This means that for high initial lags (corresponding to strong pinning regions) vortices experience an initially increasing drag after the depinning, as in the $\beta<0$ power law case. 
On the contrary,  for low initial lags only $\beta>0$ regions are sampled, see \cref{fig:RildeInterpolPicture}. 
We explore the effect of this drag model on the observed glitch size distribution with both our standard 2-component model, and also with an approximate 3-component model. 
In both cases the observed glitch distribution presents an excess of large glitches. 
This automatically implies that, given an input distribution that favors small initial lags \citep{HowittApJ2018}, the observed glitch distribution is bimodal, with a narrower peak above $\Delta\nu_{glitch}\approx 10^2 \mu Hz$ and a wider component for lower sizes.
This is qualitatively consistent with what is observed in the pulsar population \citep{fuentes17}. 
It is thus possible that both populations of glitches (i.e. the `large' and `small' glitches) originate from the same mechanism, namely the recoupling of a pinned superfluid component, once the realistic kelvin-phonon mutual friction is considered.

Furthermore, we speculate that the different size distributions observed in individual pulsars may be due to the glitch originating in different regions of pinned vorticity. For pulsars where power law distributions are observed, it is likely that the glitch originates in regions where the pinning is not strong enough to allow for large initial lags (i.e. regions in which the typical lags before unpinning are not large enough to allow the vortex to experience the kelvonic, $\beta= -3/2$, branch of the mutual friction). 
On the other hand, in pulsars that glitch quasi-periodically with a preferred size, such as the Vela or J0537-6910, it is likely that glitches occur in strong pinning regions where only kelvonic mutual friction is present. 
This would be the case in the outer core, where vortex-flux tube interactions allow for strong pinning \citep{SourieChamel2020}, but will also excite Kelvin waves on the vortices once they are free to cut through the flux tubes \citep{Ruderman1998,Link2003L}.

In conclusion we have shown that non-linear mutual friction in NS interiors leads to appreciable differences in pulsar glitch rises compared to the standard linear model. 
In particular, a non-linear drag that interpolates between the phononic and kelvonic regimes allows to explain the differences observed in the size distributions in terms of a single process and is also consistent with recently observed glitches in the Crab and Vela pulsar \citep{haskell2018corecrust}. 
To obtain constraints on the EOS, on transport parameters in the NS interior and on the glitch trigger region, however, future work should aim to include the effect of general relativity \citep{SC16,antonelli17b,gavassino2020universal}, and to make contact with microphysical calculations of interactions between vortices and ions in the NS crust \citep{seveso_etal16,Wlazlowski2016}. 
This will allow to study how structural differences  between glitching pulsars affect their glitch size distribution, and to constrain microphysical parameters in the high density interior of the star \citep{Ho2015,pizzochero17,montoli_eos_2020}.

%%%%%%%%%%%%%%%%%%%%%%%%%%%%%%%%%%%%%%%%%%%%%%%%%%%%
%  %  %  %  %  %  %  %  %  %  %  %  %  %  %  %  %  %  
%%%%%%%%%%%%%%%%%%%%%%%%%%%%%%%%%%%%%%%%%%%%%%%%%%%%

\section*{Acknowledgements}

T.C. acknowledges support from PHAROS COST Action (CA16214).
V.K., M.A. and B.H. acknowledge support from the Polish National
Science Centre grant SONATA BIS 2015/18/E/ST9/00577, P.I.: B. Haskell. 
This research was supported in part by the INT's U.S. Department of Energy grant No. DE-FG02- 00ER41132. 
The authors thank N. Andersson for reading the manuscript and useful critical comments. We also thank the anonymous referees for the constructive comments.

\subsection*{Data Availability}
The data underlying this article are available in the article and in its online supplementary material.
%%%%%%%%%%%%%%%%%%%%%%%%%%%%%%%%%%%%%%%%%%%%%%%%%%%%
%  %  %  %  %  %  %  %  %  %  %  %  %  %  %  %  %  %  
%%%%%%%%%%%%%%%%%%%%%%%%%%%%%%%%%%%%%%%%%%%%%%%%%%%%

\bibliographystyle{mnras}
\bibliography{gen_MF}

%%%%%%%%%%%%%%%%%%%%%%%%%%%%%%%%%%%%%%%%%%%%%%%%%%%%
%  %  %  %  %  %  %  %  %  %  %  %  %  %  %  %  %  %  
%%%%%%%%%%%%%%%%%%%%%%%%%%%%%%%%%%%%%%%%%%%%%%%%%%%%

\bsp
\label{lastpage}
\end{document}